# Sequences of Exact Analytical Solutions for Plane-Waves in Graded Media

Jean-Claude Krapez

*ONERA, The French Aerospace Lab, DOTA, F-13661 Salon de Provence, France*

krapez@onera.fr

Tel: +33 4 90 17 01 17, Fax: +33 4 90 17 01 09

# Sequences of Exact Analytical Solutions for Plane-Waves in Graded Media

We present a new method for building sequences of solvable profiles of the electromagnetic (EM) admittance in lossless isotropic materials with 1D graded permittivity and permeability (in particular profiles of the optical refractive-index). These solvable profiles lead to analytical closed-form expressions of the EM fields, for both TE and TM modes. The Property-and-Field Darboux Transformations method, initially developed for heat diffusion modelling, is here transposed to the Maxwell equations in the *optical-depth space*. Several examples are provided, all stemming from a constant seed-potential, which makes them based on *elementary functions* only. Solvable profiles of increasingly complex shape can be obtained by iterating the process or by assembling highly flexible canonical profiles. Their implementation for modelling optical devices like matching layers, rugate filters, Bragg gratings, chirped mirrors or 1D photonic crystals, offers an exact and cost-effective alternative to the classical approaches.

## 1. Introduction

This paper deals with the construction of exact closed-form analytical solutions of the Maxwell equations for the purpose of simulating electromagnetic (EM) plane-wave (possibly tilted) propagation in graded isotropic media where both permittivity and permeability are real-valued and present (possibly piecewise-) continuous variations in one direction. Included are the non-magnetic, i.e. optical materials, in particular rugate filters, Bragg mirrors, index matching layers, photonic crystals… Moreover, since the time-dependent harmonic equations for the electric and magnetic fields are common to other wave propagation phenomena, the method that will be described thereafter can be easily transposed to, e.g. acoustic waves and transmission lines, just to name a few other cases.

The EM-wave equations suffer from the lack of a general exact and closed-form analytical solution in the case of arbitrary permittivity and permeability distributions. Exact solutions were found only for a few particular 1D profiles as shown in textbooks [1]-[3] and therein cited references; see also [4]-[13]. The inconvenient with most solutions is that they are expressed in terms of special functions (i.e. Bessel,

Airy, Heun, Mathieu or hypergeometric functions). However, for ease and speed of computation of the solutions, especially in the perspective of solving inverse or optimization problems, the availability of closed-form solutions involving only *elementary* functions is highly desired. A few such solutions exist, however the associated permittivity or refractive index profiles involve only a small number of free-parameters: up to three in [3], [10]-[12], four in [13]. However, when aiming at modelling complex-shape refractive index profiles, parametric solutions involving a significant number of free-parameters are a key asset. Let us mention that in the quest for exact solutions for modelling 1D *periodic* optical devices (1D photonic crystals), classes of solvable profiles based on trigonometric functions were produced along with a methodology for iteratively enriching them [14]-[18].

The classical (approximate) method for modelling light waves in graded slabs consists in discretizing them in thin homogeneous layers (substantially thinner than the smallest wavelength under consideration) and to apply the well-known analytical transfer matrix approach [19]-[21]. Replacing the elementary constant profiles by smooth profiles for which an analytical transfer matrix is available would allow reducing the number of discretizing layers. As such, transfer matrices were proposed in [8] and in [9] for elementary index profiles of exponential type, resp. sinusoidal type. However, again, they require the computation of special functions, namely Bessel functions, resp. Mathieu functions.

The solution proposed in [22] is a *power series* solution in the wave-number (Spectral Parameter Power Series method - SPPS). The coefficients in the series are recursive integrals involving the refractive index profile as a weighting function. The SPPS method can cope with arbitrary permittivity profiles. High precision can be reached, however at the expense of high computing time [23].

As pinpointed in [24], the formal analogy of the Helmholtz equation of monochromatic light-waves (as expressed in the *physical-depth space*) and the stationary Schrödinger equation explains that several methods commonly applied in quantum mechanics recently found a great interest in the optical realm: the supersymmetry (SUSY) method, the Hamiltonian factorization method and the Darboux transformation (DT) method [25], [26]. Close connections can be found between them when considering the involved differential transformations. Basically, these methods allow building families of progressively more complex (solvable) potential-functions in the Schrödinger equation, along with the corresponding wave-function solutions, starting from a simple (possibly trivial) solvable potential. The DT was implemented for synthetizing passive optical waveguides and interconnects [27] and for designing active waveguides with individually tailored modal gain [28]. It was also recently applied to the synthesis of Bragg filters with a given number of resonances at desired frequencies [24]. SUSY optics was applied for tailoring the refractive index structure in the active (waveguide) layer of quantum cascade lasers [29] and for designing a partner permittivity-profile leading to the same scattering characteristics (reflection/transmission) as the original one but presenting lower contrast dielectric arrangements thus allowing to meet the technical (fabrication) specifications more easily [30]. In the latter works, the SUSY methodology was applied to the Helmholtz equation satisfied by the electric field in the TE mode when expressed in the *physical-depth space*. The same approach for the TM mode proved to be more involved. In both cases, only the permittivity (not the permeability) was assumed to be space-dependent (1D). In [31] a 2D (separable) space-dependence was assumed for the permittivity field. In [32] both permittivity and permeability presented one-dimensional space variability; a quantum mechanical analogy of the equations describing the TE and TM modes was then established for wave modes in planar waveguides. However, since they are expressed in the *physical-depth space*, these equations involve multiple permittivity/permeability derivatives which restricted the analysis to the case of a *constant* refractive index. Notice that a similarly restrictive condition, namely a constant wave-impedance, was assumed in [33] for getting solvable profiles of hyperbolic-tangent type.

Linear acoustic waves are also described by a Helmholtz equation, which explains that the Darboux transformation was also applied for solving spectral problems in acoustics [34] and for constructing solvable sound velocity profiles [35].

The method that will be described in this paper is a direct transposition to the scalar Maxwell equations of a method initially devised for modelling *heat diffusion* in graded media [36]. Advantage was taken of the fact that the Liouville transformation [37] changes the heat diffusion equation into a stationary Schrödinger equation where the potential-function corresponds to the *relative curvature* of a canonical property, namely the *square root of thermal effusivity*. Actually this also occurs with a special heat equation expressed in terms of the *heat flux density* instead of the *temperature*: in this case the potential-function amounts to be the relative curvature of the *reciprocal square root effusivity*. The joint Property & Field Darboux Transformations method (PROFIDT) was then devised for building concomitantly sequences of solvable effusivity profiles and the related temperature (or heat flux) solutions [36]. We will show that the Liouville transformation changes the scalar Maxwell equations for the electric field and for the magnetic field into two Schrödinger equations as well. Furthermore, the potential-functions have the same functional form as described before, namely they correspond to the relative curvature of a canonical EM property. Obviously, the PROFIDT method can then be applied as well which is the central objective of present paper.

The preliminary Liouville transformation amounts to change from the *physical-depth space* to the wave travel-time space, or equivalently to the *optical depth space.* Performing a preliminary Liouville transformation to the wave equation is a common procedure in inverse-scattering (see e.g. [2]). The purpose in [11] was to find new solvable profiles for the 1D acoustic wave and transmission-line problems. The authors then relied on the inverse-scattering Gel'fand-Levitan-Marchenko method, while restricting to the case where the reflection coefficient is a rational function of the wavenumber. We believe that the PROFIDT method is simpler and that it allows building higher-order solutions through a straightforward and systematic iterative procedure. The authors in [34] also performed a preliminary Liouville transformation to the (acoustic) wave equation; thereafter they developed a specific Darboux transformation formalism while renouncing to apply the "classical" Darboux transformation contrarily to what will be done with the PROFIDT method jointly on the EM-property equation and on the EM-field equation. In [35], only a *partial* Liouville transformation was performed (i.e. with no independent-variable change). The objective of the present work will be to highlight the benefit in performing a *full* Liouville transformation on both wave equations for electric and magnetic fields, followed by Darboux transformations according to the PROFIDT method.

Taking advantage of the fact that the Schrödinger potential in a Sturm-Liouville equation may be cast into the relative curvature of a canonical property and thereafter applying joint Darboux transformations is actually not new. This technique was already applied in mechanics inverse modelling for the purpose of designing isospectral elastic rods with variable cross section, acoustic horns, taut strings [38], isospectral beams governed by the Euler-Bernouilli equation [39] and families of beams presenting the same sequence of buckling loads [40]. The determination of quasi-isospectral longitudinally vibrating rods and bending rods with a prescribed set of natural frequencies was later considered in [41], resp. [42]. More recently, iterated double joint Darboux transformations were applied for characterizing a modified duct profile (locally reduced cross-section) or a damaged rod (locally reduced axial stiffness) through a matching of the altered spectrum of the acoustic wave in the duct [43], resp. the longitudinal vibration in the rod [44]. In all these mechanical applications [38]-[44], the focus was made on the eigenvalue manipulation (i.e. natural frequencies), given a specific set of boundary conditions. Furthermore, in all cases except in [44], the canonical property whose profile was under study is defined by a single mechanical property or a pair of *dependent* properties. The PROFIDT method in [36], as well as its adaptation in the present paper, was built with other objectives: solvable profiles of the canonical property are looked for, irrespective of any boundary condition; therefore, there is no consideration of eigenvalues or operator spectrum. In addition, the canonical property is defined from two *independent* space-variable properties: heat capacity and conductivity in [36], permittivity and permeability in the present paper.

Our presentation will be restricted to EM-waves in lossless materials, i.e. with real-valued permittivity and permeability. The method will be described for the general case where both properties are variable along one direction and for both TE and TM wave-polarizations. The output consists of solvable profiles of the EM *admittance* together with the related analytical closed-form solutions for the EM-fields. In the case of non-magnetic materials, the obtained profiles relate to the *tilted optical refractive index*. Two main paths will be described. The first one yields solvable profiles that are defined with enough parameters for fitting to any reasonably complex real profile. The second one lies on a class of canonical solvable profiles that present two advantages: they are highly flexible and they have an easy-to-compute transfer matrix; they will be used as elementary bricks for synthetizing admittance (or refractive-index) profiles of more sophisticated shape.

The paper is organized as follows. In section 2 we expose the equations describing the EM-wave model in the optical-depth space (Liouville transformation). In section 3 we sketch the different steps of the proposed method (adapted PROFIDT method, transfer matrices, inverse Liouville transformations). In section 4 we present the *fundamental* solutions of the Schrödinger equation (i.e. those derived from a *constant potential*). These will be used for initiating the Darboux transformations. Section 5 is devoted to the presentation of a series of examples when applying the PROFIDT method to the aforementioned fundamental (or seed) solutions. The transformed admittance or refractive index profiles and, in some cases, the corresponding reflection or transmission spectra, are shown for illustration. In section 6 we present a few applications from a family of refractive-index profiles that show a high "flexibility". Section 7 is a summary of the achievements and a comparison with the Darboux transformation when applied in the *physical-depth space* (as opposed to the *optical-depth space,* as considered in this paper). The annexes include the principle of the PROFIDT method, the related transfer-matrix construction, some tools for implementing the inverse Liouville transformation and a discussion on other wave-like phenomena to which the present PROFIDT method could apply straightforwardly.

## 2. EM plane-wave description in the Liouville space

### *2.1. Maxwell's equations for TE and TM modes*

We consider the propagation of an electromagnetic plane-wave in an inhomogeneous isotropic medium. We assume that the medium is charge-free and lossless, with real-valued and positive permittivity and permeability. The inhomogeneous medium extends from $z = 0$ to $z_1$; an incident homogeneous medium of refractive-index $n_a$ is on the left ($z < 0$) and a homogeneous substrate refractive-index $n_s$ is covering the half-space $z > z_1$, see Figure 1. In another configuration, the graded medium covers the whole half-space $z > 0$. Permittivity and permeability can only vary in $z$-direction: $\varepsilon(z)$, $\mu(z)$ (this $z$-dependence will however often be dropped in the equations). An EM plane-wave is incident from the left of the material with an angle of incidence $\phi_a$ in the $y$ - $z$ plane.

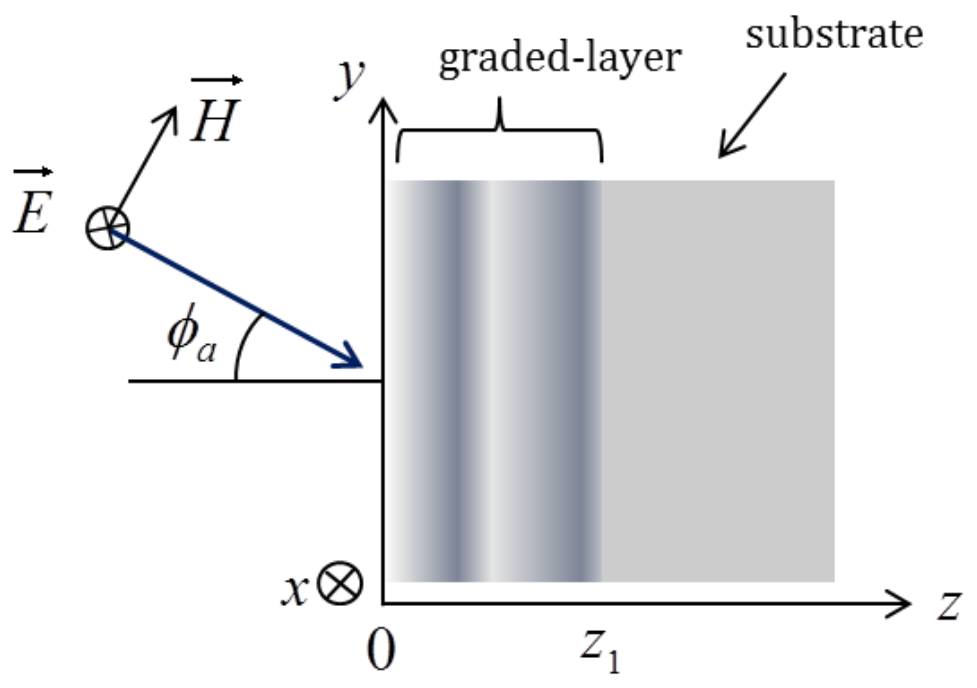

Figure 1. Considered geometry. Electromagnetic-field vectors in the case of TE-wave with incidence angle $\phi_a$.

In order to determine the EM fields, we will start by applying the formalism used in [45], [46], albeit in a condensed form that allows considering in the same time the TE (or s-) polarization and the TM (or p-) polarization. The components of the electric field and magnetic field take the periodic time-dependence form $f(z)\exp[ik_0 c(t - yI_a/c)]$ where $c$ and $k_0$ are the speed of light and the wavenumber in vacuum, whereas $I_a$ is the Snell-Descartes invariant defined by:

$$I_a = n_a \sin\phi_a = n(z)\sin\phi(z) = \sqrt{\varepsilon_r(z)\mu_r(z)}\sin\phi(z) = n_s \sin\phi_s \ , \qquad (1)$$

where $n_{a,s}$ and $\phi_{a,s}$ are the refractive-index and the incidence angle in the homogeneous media at left (*a* for air, say) and at right (*s* for substrate) of the graded material. $n$, $\varepsilon_r$, $\mu_r$ and $\phi$ are the local refractive index, relative permittivity, relative permeability and incidence angle (all four parameters may depend on $z$). The Maxwell's equations lead to the following system of coupled first-order ordinary differential equations (ODE) for the tangential components (i.e. in the $x$ - $y$ plane):

$$\begin{cases} \dfrac{dE}{dz} = -mik_0 c\mu \cos^{1-m}\phi\, H & (2) \\ \dfrac{dH}{dz} = -mik_0 c\varepsilon \cos^{1+m}\phi\, E, & (3) \end{cases}$$

where $E$ designates $E_x$ or $E_y$ depending on whether the polarization mode is TE or TM; symmetrically, $H$ designates respectively $H_y$ or $H_x$ (in the sequel, this will be called the "*XY-convention for* $E$ *and* $H$"). Furthermore, $m = +1$ for the TE mode and $m = -1$ for the TM mode. By substituting $dH/dz$ from eq. (3) into the z-derivative of eq. (2), resp. $dE/dz$ from eq. (2) into the z-derivative of eq. (3), we get the following *variable-coefficient* 1D Helmholtz equations for $E$, resp. for $H$:

$$\langle E \rangle: \quad -k_0{}^2 E = \frac{1}{c\varepsilon \cos^{1+m}\phi} \frac{d}{dz}\left( \frac{1}{c\mu \cos^{1-m}\phi} \frac{dE}{dz} \right). \tag{4}$$

$$\langle H \rangle: \quad -k_0{}^2 H = \frac{1}{c\mu \cos^{1-m}\phi} \frac{d}{dz}\left( \frac{1}{c\varepsilon \cos^{1+m}\phi} \frac{dH}{dz} \right). \tag{5}$$

In the sequel, we will use the symbol $\langle E \rangle$, resp. $\langle H \rangle$, for designating an equation that is related to the electric field component, resp. to the magnetic field component, as well as for designating the results derived therefrom.

### *2.2. Liouville transformation*

The equations (4) and (5) are 2$^{nd}$ order ODEs of Sturm-Liouville type with *variable coefficients*. The Liouville transformation allows changing such equations into ODEs in the Liouville *normal form* [37], which actually corresponds to the now so-called *stationary Schrödinger equation*. It is now a common tool in the realm of inverse scattering [2], [11]. It was also applied in [45], [46] and to some extend in [3], [10] although not being mentioned. The Liouville transformation consists both of an independent-variable change and of a dependent-variable change. The space variable $z$ is changed into the *tilted* (or *effective*) optical-thickness (or optical-depth) $\xi$, taken from 0 to $z$:

$$z \rightarrow \xi = c\int_0^z \sqrt{\varepsilon(u)\mu(u)}\cos\phi(u)du. \tag{6}$$

The variable $\xi/c$ corresponds to the wave travel-time and $k_0\xi$ corresponds to the so-called optical phase thickness. Since the argument of the integral is positive, the $z \rightarrow \xi$ mapping is one-to-one which implies that the inverse mapping exists. The dependent-variable change is defined differently for $E$ and for $H$:

$$E \rightarrow \overline{E} \equiv E\eta^{1/2} \tag{7}$$

$$H \rightarrow \overline{H} \equiv H\eta^{-1/2}, \tag{8}$$

where $\eta$ is the *tilted* (or *effective*) admittance which is defined by:

$$\eta \equiv \sqrt{\varepsilon/\mu}\cos^m\phi = \sqrt{\varepsilon/\mu}\left(1 - \frac{I_a^2}{c^2\varepsilon\mu}\right)^{m/2}, \tag{9}$$

where we used eq. (1) for eliminating the incident angle $\phi(\xi)$. The ODE (2) and (3) are then transformed into the coupled equations:

$$\begin{cases} E' = -mik_0\eta^{-1}H & (10) \\ H' = -mik_0\eta E, & (11) \end{cases}$$

where (and from now on) the primes denote derivation with respect to the tilted optical-depth $\xi$. On the other side, the equations (4) and (5) are transformed into two equations in the *Liouville normal form*:

$$\langle E \rangle: \qquad \bar{E}'' = \left( \eta^{1/2\,''} / \eta^{1/2} - k_0^2 \right) \bar{E} \tag{12}$$

$$\langle H \rangle: \qquad \bar{H}'' = \left( \eta^{-1/2\,''} / \eta^{-1/2} - k_0^2 \right) \bar{H} \,. \tag{13}$$

Notice that in [46], the first term in bracket was expressed in terms of $\eta$, not $\eta^{1/2}$ nor $\eta^{-1/2}$, which led to more complicated expressions and required introducing an approximation. Actually, the pair of equations (12) and (13) can be considered as a generalization to both fields $E$ and $H$ of the result obtained for only one modal-field in [10]. Transformations like those in eq. (7)-

(8) were also applied in [3] but the space variable was changed into the optical-thickness, not the *tilted* optical-thickness, which led to more complicated expressions. Interestingly, eq. (12) and (13) look quite simple and perfectly symmetrical. This compactness highlights the benefit of moving from the physical-depth space to the *tilted* optical-depth space. What is more, no approximation was introduced, contrarily to the coupled-wave theory for local-periodic gratings which is restricted to shallow up to moderate gratings (see e.g. [24], [47]).

It may seem odd to develop in parallel the 2nd order ODEs for *both* $E$ and $H$ since once one of these fields is known, the other one can be inferred directly from eq. (2) or eq. (3). As it will soon become clear, considering *concomitantly* both fields $E$ and $H$ will give rise to *twice more solutions*. Indeed, since eq. (12) and eq. (13) are identical, except the admittance is changed into its reciprocal, we can state that if a given admittance profile is *solvable* then the same is true for its *reciprocal*.

### *2.3. Comparison with the heat-diffusion problem*

Eq. (12) and (13) actually take the form of a time-independent Schrödinger equation $\left( -\frac{d}{d\xi^2} + V \right) \psi = \mathbb{E} \psi$ :

the *relative curvature* of $\eta^{1/2}$ (resp. $\eta^{-1/2}$) vs. the tilted optical-thickness $\xi$ plays the role of the potential-function $V$ and $k_0^2$ plays the role of the state energy $\mathbb{E}$. Equations (7), (8), (12) and (13) present strictly the same form as those obtained in [36] in the context of heat diffusion in graded media. In both cases the initial transfer equations have been transformed into Schrödinger equations where the potential-function expresses as the *relative curvature* of a specific parameter. The analogy is actually as follows:

$$\left(E, H, \eta, -k_0{}^2\right) \leftrightarrow (\theta, \phi, b, p) \tag{14}$$

where $\theta$ and $\phi$ are the (time-)Laplace-Fourier transforms of the temperature, resp. of the heat flux, $b$ is the thermal effusivity profile and $p$ is the (real-valued, positive) Laplace variable or the (pure imaginary complex) Fourier variable, depending on the dynamic evolution under consideration, either transient or steady periodic. Note that in [36], the Liouville (independent-) variable $\xi$ stands for the square-root of the heat diffusion-time along the [0, $z$ ] interval.

Obviously the technique that was developed in [36] for building *solvable profiles of the thermal effusivity* $b(\xi)$ should apply equally well for building *solvable profiles of the EM admittance* $\eta(\xi)$. Even more, the functions describing solvable profiles of thermal *effusivity* will prove to describe solvable profiles of EM *admittance* as well (up to a scaling parameter). However, as it will be shown later, the solvable profiles of the *refractive-index* may be of different form, depending on the polarization and the incidence angle. Furthermore, the inverse-Liouville transformation itself will introduce other peculiarities. In addition, because of the different nature of the "eigenvalue-parameter": $p$ (real-valued positive or pure imaginary) as compared to $-k_0{}^2$ (real-valued negative), substantial differences will appear in the field-solution $\psi(\xi, k_0)$ construction (this comes naturally from the different nature of the transfer processes: diffusion as opposed to wave propagation) as well as in the inferred scattering properties (e.g. temperature contrast vs. spectral reflectance).

As a consequence, there is no need repeating the description of the method that was already published in [36]. We will merely provide the outline of the process. The focus will be done on the salient features related to the present EM-wave problem.

## 3. Outline of the resolution method

### *3.1. Formulation into a system of two homologous Schrödinger equations*

Each of the two Schrödinger equations in eq. (12) and eq. (13) can be cast into two *homologous* Schrödinger equations, which means two equations sharing the *same potential-function* $V(\xi)$:

$$\begin{cases} \psi'' = \left(V(\xi) - k_0^2\right)\psi & (15) \\ s'' = V(\xi)s, & (16) \end{cases}$$

with the following correspondences between, on one side, the new *field (or wave-function)* $\psi(\xi, k_0)$ and the *transformed EM-admittance* $s(\xi)$, and, on the other side, the former electric and magnetic fields, together with the tilted admittance $\eta$:

$$\langle E \rangle : \quad \begin{cases} \psi \leftrightarrow E\eta^{1/2} & (17) \\ s \leftrightarrow \eta^{1/2} & (18) \end{cases}$$

$$\langle H \rangle : \quad \begin{cases} \psi \leftrightarrow H\eta^{-1/2} & (19) \\ s \leftrightarrow \eta^{-1/2} & (20) \end{cases}$$

The problem is now to find "solvable" potential functions $V(\xi)$, which, in the present context, means finding potentials that lead to closed-form analytic solutions for both $\psi(\xi, k_0)$ in eq. (15) and $s(\xi)$ in eq. (16). Since both eq. (15) and eq. (16) are homogeneous 2$^{\text{nd}}$ order ODE, their generic solutions can be expressed as linear combinations (LC) of two independent solutions: $K(\xi, k_0)$ and $P(\xi, k_0)$ for the former, resp. $B(\xi)$ and $D(\xi)$ for the latter. This will be expressed through the symbol $\propto$, while omitting the constant coefficients:

$$\psi(\xi,k_0)\propto\begin{bmatrix}K(\xi,k_0)\\P(\xi,k_0)\end{bmatrix};\quad s(\xi)\propto\begin{bmatrix}B(\xi)\\D(\xi)\end{bmatrix}. \tag{21}$$

Selected classes of linear independent solutions (LIS) will be presented in §4-5 and summarized in Table 1. Because of the already mentioned similarities with the heat diffusion problem, this table share several columns with the corresponding Table 1 in [36]. In particular, column 7 which reports functions $B(\xi)$ and $D(\xi)$ is substantially the same in both tables. On the contrary, column 9 which reports the independent solutions $K(\xi,k_0)$ and $P(\xi,k_0)$ is different: the basis functions in [36] are hyperbolic functions whereas in present Table 1 they are exponential functions of positive (or negative) argument describing forward (resp. backward) propagating waves.

### *3.2. The trivial case of a constant potential : fundamental solutions*

Trivial solutions to eq. (15)-(16) are obtained when setting the potential to a constant value: $V(\xi)=\beta$. We will call them the *fundamental solutions*. Their properties will be described in §4. The results regarding $s(\xi)$ and $\psi(\xi,k_0)$ are summarized as cases c1, c6 and c9 in Table 1.

### *3.3. Beyond the fundamental solutions: application of the Darboux transformation*

The Darboux transformation (DT) allows building sequences of solvable (Schrödinger) potentials, starting from a (possibly trivial) solvable potential. More precisely, if one knows the general solution of eq. (15) for a given potential $V(\xi)$ and for any value of the "eigenvalue" $-k_0^{\,2}$ , the DT allows building a new potential together with the corresponding general solution for any value of the "eigenvalue". The new potential and the new solutions are defined with (up to) two parameters more than at the previous stage. The process can be iterated, hence enriching the potential and the solutions.

The outstanding point in eq. (15) and eq. (16) is that the transformed property $s(\xi)$ satisfies a Schrödinger equation with the same potential function as do the transformed field ($\psi\equiv E\eta^{1/2}$ or $H\eta^{-1/2}$), albeit with a vanishing parameter, i.e. $k_0$ =0. A direct consequence is that performing a DT on eq. (15)

will simultaneously provide a new set of solutions for the field $\psi(\xi, k_0)$ and a new set of property functions $s(\xi)$. This combined method yielding concomitantly new solvable profiles of the *property* $s(\xi)$ *and* the related new *field* functions $\psi$ was named the PROFIDT method in [36] (i.e. PROperty and FIeld Darboux Transformations). When applied recursively, it allows building more and more complex $\eta(\xi)$ profiles together with the related $\psi(\xi, k_0)$ solutions. In addition, starting from a *constant potential* with the associated *fundamental solutions* as seed solutions has the advantage of generating solutions that are based on *elementary functions* only. The PROFIDT method was fully described in [36]; we briefly review its essence in Annex A for the sake of completeness.

The fact that the leading property profile and the field function satisfy two homologous equations was already noticed by former authors in the context of acoustics in [11] and in [34], although not explored further. Actually, the authors in [34] developed a specific Darboux transformation formalism for a modified equation $\psi'' = U(\xi)\psi' - k_0^2\psi$ involving a potential-function of the form $U(\xi) = -\eta'/\eta$ while renouncing to apply the classical Darboux transformation directly on eq. (15) and eq. (16), as is actually done in the PROFIDT method.

Joint Darboux transformations on a system like eq. (15)-(16), as obtained from an equation of the same form as eq. (12), can also be found in a series of works on inverse modelling in mechanics [38]-[44]. As explained in the introduction, the purpose was then "eigenvalue engineering" (i.e. natural-frequency manipulation). By contrast, our aim is to find out new solvable profiles for the admittance $\eta(\xi)$ (irrespective of boundary conditions) and to build the corresponding transfer matrices in view of the computation of the scattering properties of composite profiles.

### *3.4. Inverse Liouville transformation: from $\{s, \psi\}$ to $\{\eta, E, H\}$*

Before providing sequences of solutions to eq. (15) and eq. (16), we have to describe some tools for later handling them. The interesting point is that one can get, from *one* given set of solutions $\{s, \psi\}$ of eq. (15)-(16), *two* sets of tilted admittance and EM-fields: either $\{\eta, E\}$ by referring to the $\langle E \rangle$-form transformation in eq. (17)-(18), or $\{\eta, H\}$ by referring to the $\langle H\rangle$-form transformation in eq. (19)-(20). For getting the missing EM-field ($H$ in the first case and $E$ in the second one), one still has to use eq. (2), resp. eq. (3) while considering the admittance definition in eq. (9). The operations are summarized in the following expressions, common to both polarizations (the mode-coefficient $m$ and the XY-convention for $E$ and $H$ allow switching from one mode to the other):

$$\langle E\rangle : \begin{cases} \eta = s^2 & (22) \\ \begin{cases} E = \psi s^{-1} & (23) \\ H = mik_0^{-1} W(s,\psi) & (24) \end{cases} \end{cases}$$

$$\langle H\rangle : \begin{cases} \eta = s^{-2} & (25) \\ \begin{cases} E = mik_0^{-1} W(s,\psi) & (26) \\ H = \psi s^{-1} & (27) \end{cases} \end{cases}$$

where $W(s,\psi)$ is the Wronskian determinant of $s$ and $\psi$, namely $s\psi' - s'\psi$.

Among all solutions generated by the linear combination (LC) of $B(\xi)$ and $D(\xi)$, only those leading to physically acceptable admittance profiles $\eta(\xi)$ (through eq. (22) or eq. (25)) should be retained, which means, for the EM "standard" case, positive, non-vanishing and bounded values over the considered domain. In the case of a finite layer, the admittance profiles will be described, based on the (relative) admittance values at the layer edges $\eta_0$ and $\eta_1$, by :

$$\eta^{\pm 1/2}(\xi) = \eta_0^{\pm 1/2} \frac{B(\xi)D_1 - B_1 D(\xi)}{B_0 D_1 - B_1 D_0} + \eta_1^{\pm 1/2} \frac{B_0 D(\xi) - B(\xi) D_0}{B_0 D_1 - B_1 D_0}, \quad (28)$$

where the indices 0 and 1 indicate that the functions are expressed at the left, resp. right edge of the considered layer, i.e. at $\xi = 0$ and $\xi = \xi_1$. Notice that in this paper the shorthand notation $\pm 1/2$ for an exponent should be understood as follows: the positive exponent $+1/2$ refers to the profiles inferred from the $\langle E \rangle$-form equation whereas the negative exponent $-1/2$ refers to those inferred from the $\langle H \rangle$-form equation.

### *3.5. Transfer-matrices related to solutions expressed in the Liouville space*

From the EM fields obtained from eq. (23)-(24) or from eq.

(26)-(27) it is then an easy (although cumbersome) task to build the transfer matrices. The transfer matrix is aimed at providing a link between the vector of EM tangential components $[E_1, H_1]^t$ at abscissa $\xi = \xi_1$ and the same vector $[E_0, H_0]^t$ at another abscissa $\xi = \xi_0$:

$$\begin{bmatrix} E_0 \\ H_0 \end{bmatrix} = \mathbf{M} \begin{bmatrix} E_1 \\ H_1 \end{bmatrix}; \quad \mathbf{M} = \begin{cases} \mathbf{M}_{\langle E \rangle} \\ \mathbf{M}_{\langle H \rangle} \end{cases} \tag{29}$$

The analytical expressions of the four terms of the transfer matrix $\mathbf{M}_{\langle E \rangle}$ are presented in Annex B. When comparing eq. (23)-(24) on one side and eq.

(26)-(27) on the other side, it is easy to notice that the matrix corresponding to the $\langle H \rangle$-form solution, i.e. $\mathbf{M}_{\langle H \rangle}$, is simply obtained by a two-step circular permutation of the previous matrix $\mathbf{M}_{\langle E \rangle}$:

$$\mathbf{M}_{\langle E \rangle} = \begin{bmatrix} A & B \\ C & D \end{bmatrix} \Rightarrow \mathbf{M}_{\langle H \rangle} = \begin{bmatrix} D & C \\ B & A \end{bmatrix}. \tag{30}$$

By expressing the continuity of the tangential components of the EM-field at each interface of a *multilayer* made of graded layers, the tangential components of the EM-field at both ends of said multilayer are then naturally related through a transfer matrix resulting from the product of the matrices pertaining to each layer. Thereafter, based on the knowledge of the admittances of the front material $\eta_a$ and of the substrate material $\eta_s$, the reflectance and the transmittance of an assembly of graded layers are obtained according to the same procedure as for uniform layers [19]. Equivalently they can be obtained through a recursive calculation of the surface admittance, starting from the back face (see Annex B).

### *3.6. Downscaling from $\eta(\xi)$ to the original properties $\varepsilon_r(\xi)$, $\mu_r(\xi)$, $n^*(\xi)$ and $n(\xi)$*

A solvable profile of *tilted admittance* $\eta(\xi)$ (as determined from eq. (22) or from (25)), is directly converted into a profile of the *relative tilted admittance* $\eta_r(\xi)$ by $\eta_r(\xi)=\eta(\xi)/y_0$ where $y_0=\sqrt{\varepsilon_0/\mu_0}$ is the free-space admittance, $\varepsilon_0$, $\mu_0$ are the permittivity and the permeability of vacuum. It is clear from eq. (9) that any tilted admittance profile $\eta(\xi)$ may be converted into an *infinite set* of coupled permittivity $\varepsilon(\xi)$ and permeability $\mu(\xi)$ profiles. As a consequence, it may also be converted into an infinite set of solvable refractive-index profiles $n(\xi)\equiv\sqrt{\varepsilon_r(\xi)\mu_r(\xi)}$ and thereafter an infinite set of pseudo-index (or tilted index) profiles, which are defined according to $n^*(\xi)\equiv n(\xi)\cos^m\phi(\xi)$ [19].

Evidently, for going further, ancillary information is required: it may happen that one of the two profiles, permittivity or permeability is already specified or that these two properties are bound through a constitutive relation, e.g. $\mu=h(\varepsilon)$. Yet, we won't explore further this vast potential for the generation of solvable EM-profiles which could be of interest for the design of metamaterials. Instead, each time an admittance $\eta(\xi)$ has to be expressed in terms of other EM properties (permittivity or refractive-index), or when an inverse Liouville transform has to be performed, we will assume that the medium is *non-magnetic*, i.e. $\mu_r(\xi)=1$, which anyway corresponds to a broad range of optical materials. With this restriction in mind, Table 2 reports, for both polarization modes, the relations that provide $n^*(\xi)$, $\varepsilon_r(\xi)$ and $n(\xi)$ from the *relative* admittance profile $\eta_r(\xi)$. Notice that the pseudo-index profile $n^*(\xi)$ coincides with the *relative* admittance profile $\eta_r(\xi)$. Table 2 also shows that, for the TM mode, there are two possible solutions regarding the inferred profiles $\varepsilon_r(\xi)$ and $n(\xi)$. This comes from the $\pm$ sign in front of the square root term. A close analysis reveals that the "+" solution (resp. the "-" solution) is associated with incidence angle values $\phi(\xi)$ that are lower than 45° (resp. higher). In the end, by adding up all the TE-mode and TM-mode solutions, each set of solutions $\{s(\xi),\psi(\xi,k_0)\}$ gives rise to *six* sets of solutions $\{n(\xi),E(\xi,k_0),H(\xi,k_0)\}$, without mentioning the innumerous profiles obtained in the TM-mode case by mixing sub-elements obtained by invoking either the “+” solution or the “-” solution as described in the rightmost column of Table 2.

Regarding the next graphical results about the solvable profiles, any plot of the relative tilted admittance $\eta(\xi)/\eta_0$ (valid for both TE and TM modes and for any incidence) also describes the relative tilted index $n^*(\xi)/n_0^*$ (in the case of non-magnetic materials, $\mu_r(\xi)$=1) and the relative refractive index $n(\xi)/n_0$ (when further assuming normal incidence).

### *3.7. Inverse Liouville transformation of the independent variable: from $\xi$-space back to $z$-space*

For expressing the obtained profiles back in the original $z$-space, we need, in the general case and according to eq. (6), to compute the integral:

$$z(\xi)=\int_0^{\xi}\left(\sqrt{\varepsilon_r(u)\mu_r(u)}\cos\phi(u)\right)^{-1}du=\int_0^{\xi}\left(\varepsilon_r(u)\mu_r(u)-I_a^2\right)^{-1/2}du\,, \tag{31}$$

which, again, shows a multiplicity of results. When restricting, as before, to *non-magnetic* materials, this reduces to:

$$z(\xi)=\begin{cases}\int_0^{\xi}\eta_r^{-1}(u)du & TE \quad (32)\\ 2\int_0^{\xi}\eta_r^{-1}(u)\left(1\pm\sqrt{1-4I_a^2\eta_r^{-2}(u)}\right)^{-1}du & TM. \quad (33)\end{cases}$$

The detailed treatment of these integrals is moved to Annex C. Yet, a noticeable result is that the first integral (TE-mode case) can be changed into an analytical closed-form expression by implementing a procedure that was devised in [36], based on the work presented in [48]-[50] (see Annex C). Unfortunately, no such simplification could be found for the second integral, i.e. for the TM-mode case.

Except for very simple expressions of $\eta_r(\xi)$, this integral is much likely to be evaluated numerically. For this reason, for later illustrations of the inverse Liouville-transform results, we will systematically assume $\mu_r = 1$ and consider the TE-mode only ("optical-TE" case).

## 4. Fundamental solutions (seed-solutions)

The *fundamental solutions*, namely those obtained with a constant potential $V(\xi) = \beta$ are a generalization to *arbitrary incidence* of the solutions presented in [10] for normal incidence only. In contrast to the method adopted in [10], we first express the admittance profile solutions in the $\xi$-space for both $\langle E \rangle$-form and $\langle H \rangle$-form equations. The refractive-index profile $n(\xi)$ can then be obtained from the expressions in Table 2, depending on the polarization mode.

### *4.1. Zero potential:* $V(\xi) = \beta = 0$

The generic wave-function $\psi(\xi, k_0)$ is given by the LC:

$$\psi(\xi, k_0) \propto \begin{bmatrix} K(\xi, k_0) \\ P(\xi, k_0) \end{bmatrix} \equiv \begin{bmatrix} \exp(ik_0\xi) \\ \exp(-ik_0\xi) \end{bmatrix} \equiv \begin{bmatrix} C^{(0)} \\ S^{(0)} \end{bmatrix}, \tag{34}$$

whereas the joint transformed-admittance solutions $s(\xi)$ are the linear functions of the optical-depth $\xi$:

$$s(\xi) \equiv \eta^{\pm 1/2}(\xi) \propto \begin{bmatrix} B(\xi) \\ D(\xi) \end{bmatrix} \equiv \begin{bmatrix} 1 \\ \xi \end{bmatrix}. \tag{35}$$

These fundamental solutions are reported as case c1 in Table 1. From eq. (35) we get two exact solutions corresponding to the following tilted admittance profiles:

$$\langle E \rangle: \quad \eta(\xi) = \eta_0 (1 + \alpha\xi)^2 \tag{36}$$

$$\langle H\rangle: \quad \eta(\xi)=\eta_0(1+\alpha\xi)^{-2}\ , \tag{37}$$

where $\eta_0$ is the admittance at the origin and $\alpha$ is some constant.

Let us now examine the $\langle E\rangle$-form solution (eq. (36)). After performing the inverse Liouville transformation, we get the following relation between the independent variables $\xi$ and $z$ ("optical-TE" case):

$$1+\alpha\xi=(1-\kappa z)^{-1}; \quad \kappa=\alpha n_0^{\ *}\ . \tag{38}$$

Substituting it back into the admittance profile in eq. (36), and applying the rule described in eq. (C.1) yields the index profile:

$$n(z)=\sqrt{\left(n_0^{\ 2}-I_a^2\right)(1-\kappa z)^{-4}+I_a^2}\ ; \quad \kappa=\alpha\sqrt{n_0^{\ 2}-I_a^2}\ , \tag{39}$$

(at normal incidence, it reduces to the profile described in [10]: $n(z)=n_0(1-\kappa z)^{-2}$ with $\kappa=\alpha n_0$).

Applying the same methodology with the admittance profile issued from the $\langle H\rangle$-form solution (eq. (37)), leads to the following relation between the independent variables $\xi$ and $z$:

$$1+\alpha\xi=(1+\kappa z)^{1/3}; \quad \kappa=3\alpha n_0^{\ *}\ , \tag{40}$$

and the following index profile:

$$n(z)=\sqrt{\left(n_0^{\ 2}-I_a^2\right)(1+\kappa z)^{-4/3}+I_a^2}\ ; \quad \kappa=3\alpha\sqrt{n_0^{\ 2}-I_a^2}\ , \tag{41}$$

(at normal incidence, it reduces to the profile described in [10]: $n(z)=n_0(1+\kappa z)^{-2/3}$ with $\kappa=3\alpha n_0$).

The fundamental solutions presented so far (when restricted to the half-line $z > 0$) were once obtained in the context of acoustic waves modelling by implementing the Gel'fand-Levitan inverse-scattering approach [11] (1st and 2nd cases, as inferred from a reflection coefficient defined by a "one-pole" rational function). In contrast to the Gel'fand-Levitan method, the procedure implemented in present paper is extremely simple.

### *4.2. Constant, non-zero potential* $V(\xi) = \beta \neq 0$

The general solution for the wave-function $\psi(\xi, k_0)$ is now:

$$\psi(\xi, k_0) \propto \begin{bmatrix} K(\xi, k_0) \\ P(\xi, k_0) \end{bmatrix} \equiv \begin{bmatrix} \exp\left(i\sqrt{k_0^{\ 2} - \beta}\xi\right) \\ \exp\left(-i\sqrt{k_0^{\ 2} - \beta}\xi\right) \end{bmatrix} \equiv \begin{bmatrix} C^{(\beta)}(\xi, \omega) \\ S^{(\beta)}(\xi, \omega) \end{bmatrix}. \tag{42}$$

Candidates for the admittance profile are, depending on the sign of $\beta$:

$$\eta^{\pm 1/2}(\xi) \propto \begin{bmatrix} B(\xi) \\ D(\xi) \end{bmatrix} \equiv \begin{cases} \begin{bmatrix} \cosh(\xi/\xi_c) \\ \sinh(\xi/\xi_c) \end{bmatrix} & \text{if } \beta > 0 \quad (43) \\ \begin{bmatrix} \cos(\xi/\xi_c) \\ \sin(\xi/\xi_c) \end{bmatrix} & \text{if } \beta < 0. \quad (44) \end{cases}$$

We introduced $\xi_c \equiv 1\big/\sqrt{|\beta|}$ which can be regarded as the *characteristic* optical-thickness of the profile element. These fundamental solutions are reported as cases c6 and c9 in Table 1. The constant parameters $A_B$ and $A_D$ in the LC definition of $n^{*\pm 1/2}(\xi)$ that are necessary for the inverse-Liouville transformation in eq. (C.6)-(C.12) in Annex C are given by:

$$\begin{cases} A_B = n_0^{\ *\pm 1/2} \\ A_D = \left(n_1^{\ *\pm 1/2} - n_0^{\ *\pm 1/2} B_1\right)\big/ D_1. \end{cases} \tag{45}$$

In the following, we will use the non-dimensional variables: $\hat{\xi} = \xi/\xi_c$ , $\hat{z} = zn_0^*/\xi_c$ and $\kappa = A_D/A_B$ . The admittance profiles generated by the $\langle E\rangle$-form equation (i.e. eq. (43) or (44) with exponent +1/2) lead to an explicit $n^*(\hat{z})$ relation. Indeed, we first note that $B(\xi)^2 - \mathrm{sgn}(\beta)D(\xi)^2 = 1$, which will be used for defining the function $f_{BD}$ in eq. (C.11).

Next, considering eq. (C.12) together with eq. (45) yields an explicit definition of the pseudo-index profile vs. $z$ (via $\hat{z}$ ):

$$n^*(\hat{z}) = n_0^*\left[(\kappa\hat{z} - 1)^2 - \mathrm{sgn}(\beta)\hat{z}^2\right]^{-1}. \tag{46}$$

The refractive index is finally obtained by applying the appropriate relation in Table 2. We can notice from eq. (46) that the inverse of the pseudo-index takes the simple form of a 2nd order polynomial in $z$ (this polynomial form was obtained in [10] for the particular case of normal incidence).

Unfortunately, the admittance profiles generated by the $\langle H\rangle$-form equation (i.e. eq. (43) or (44) with exponent -1/2) don't lead to an explicit expression for the pseudoindex profiles in terms of position $z$ . We have to use eq. (C.5) followed by eq. (C.8) for getting the following *implicit* relation between $z$ and $\xi$ (via $\hat{z}$ and $\hat{\xi}$ ):

$$4\hat{z} = D(2\hat{\xi}) + 2\hat{\xi} + \mathrm{sgn}(\beta)\kappa^2\left(D(2\hat{\xi}) - 2\hat{\xi}\right) + 4\kappa D^2(\hat{\xi}). \tag{47}$$

In Figure 2 we plotted a few examples from this class of "fundamental" profiles. While increasing the constant potential $\beta$ from negative to positive values, the $\langle E\rangle$-form profiles (in black) change from concave-downward to convex whereas the $\langle H\rangle$ -form profiles (in red) change from convex to concave-downward.

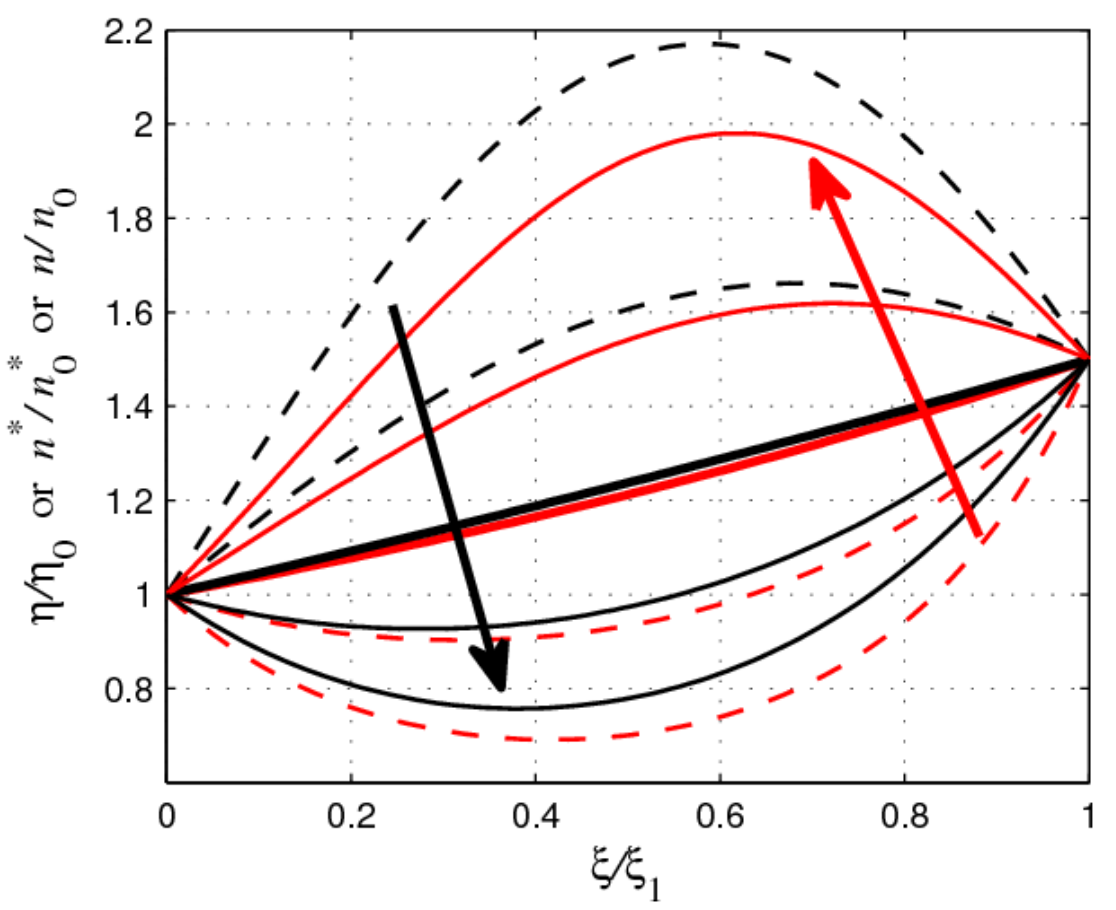

Figure 2. Solvable profiles corresponding to the fundamental solutions in eq. (35) ($\beta = 0$, bold lines), eq. (43) ($\beta > 0$, plain lines), and eq. (44) ($\beta < 0$, dashed lines). Profiles originating from the $\langle E \rangle$-form equation are in black, those originating from the $\langle H \rangle$-form are in red. The normalized parameter $\beta \xi_1^2$ is changed from -2 to +2 by steps of unity according to the arrows. The current optical-thickness $\xi$ is normalized by $\xi_1$, i.e. the optical-thickness of the graded layer. Each of the curves describes at the same time the tilted admittance $\eta(\xi)/\eta_0$ (valid for both TE and TM modes and for any incidence), the tilted index $n^*(\xi)/n_0^*$ (in the case of non-magnetic materials, $\mu_r(\xi)$=1) and the refractive index $n(\xi)/n_0$ (when further assuming normal incidence). All are normalized by the corresponding left-boundary value. Particular case of a 1.5-factor increase from left to right boundary.

In Figure 3 we plotted the same profiles as in Figure 2 but vs. the physical (normalized) depth $z/z_1$. For this representation, we took into account the inverse Liouville transform results in eq. (39) and (46) for the $\langle E \rangle$-form solutions (resp. in eq. (41) and (47) for the $\langle H \rangle$-form solutions).

In Figure 4 we simply changed from normal-incidence to an incidence of 70°. We considered here the TE-wave case only. The profound change of aspect from Figure 3 to Figure 4 highlights the fact that the solvable profiles obtained with the present method for the refractive index and for the permittivity (but not for the admittance nor for the pseudo-index) are *incidence-dependent*.

Let us now consider the configuration in Figure 1 while assuming index matching at the boundaries, i.e. $n_0 = n_a$ at the front side and $n_1 = n_s$ at the backside of the graded layer. The spectral reflectance of the ten solvable profiles in Figure 2 is reported in Figure 5. The wavelength of the EM-wave, $\lambda$, is normalized by the graded layer optical-thickness $\xi_1$. Let us mention that the scattering property computations as presented in Figure 5 for the present solvable profiles, as well as in Figure 7, Figure 9, Figure 11 and Figure 15 for the next ones, were systematically validated with the classical (i.e. piecewise uniform) transfer-matrix method.

The Fresnel reflectance value, namely 0.04 in the present case of a 50% increase of admittance from the incidence medium to the substrate is well retrieved for asymptotically large wavelength values. On the other side, the reflectance vanishes for an asymptotically short wavelength because of the continuity of admittance from the incidence medium down to the substrate. For intermediate wavelength values, i.e. for $\lambda/\xi_1$ between about 0.5 and 20, the reflectance shows variations, and possibly an overshoot, that depend on the curvature of the admittance profile.

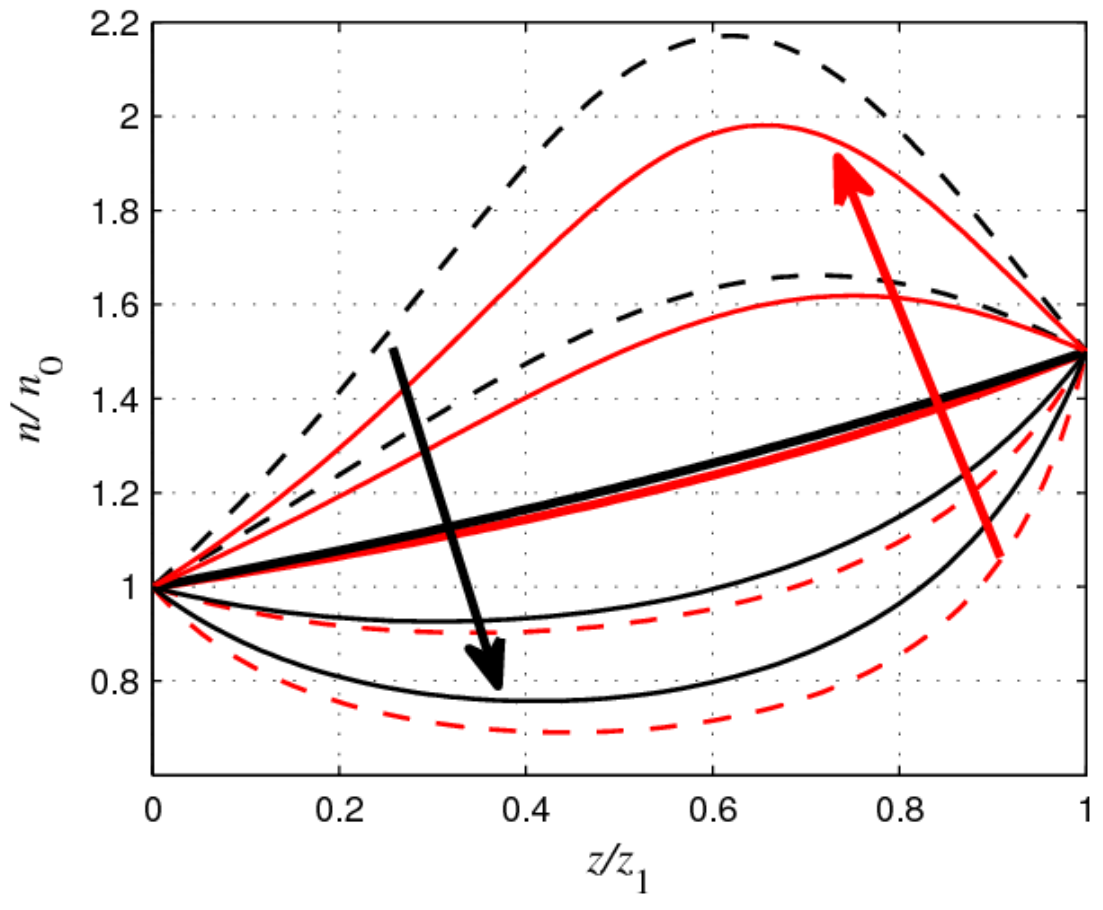

Figure 3. Solvable profiles of the refractive index expressed in terms of the relative depth $z/z_1$ where $z_1$ is the layer thickness (normal incidence). For the color and line-type significance, see Figure 2.

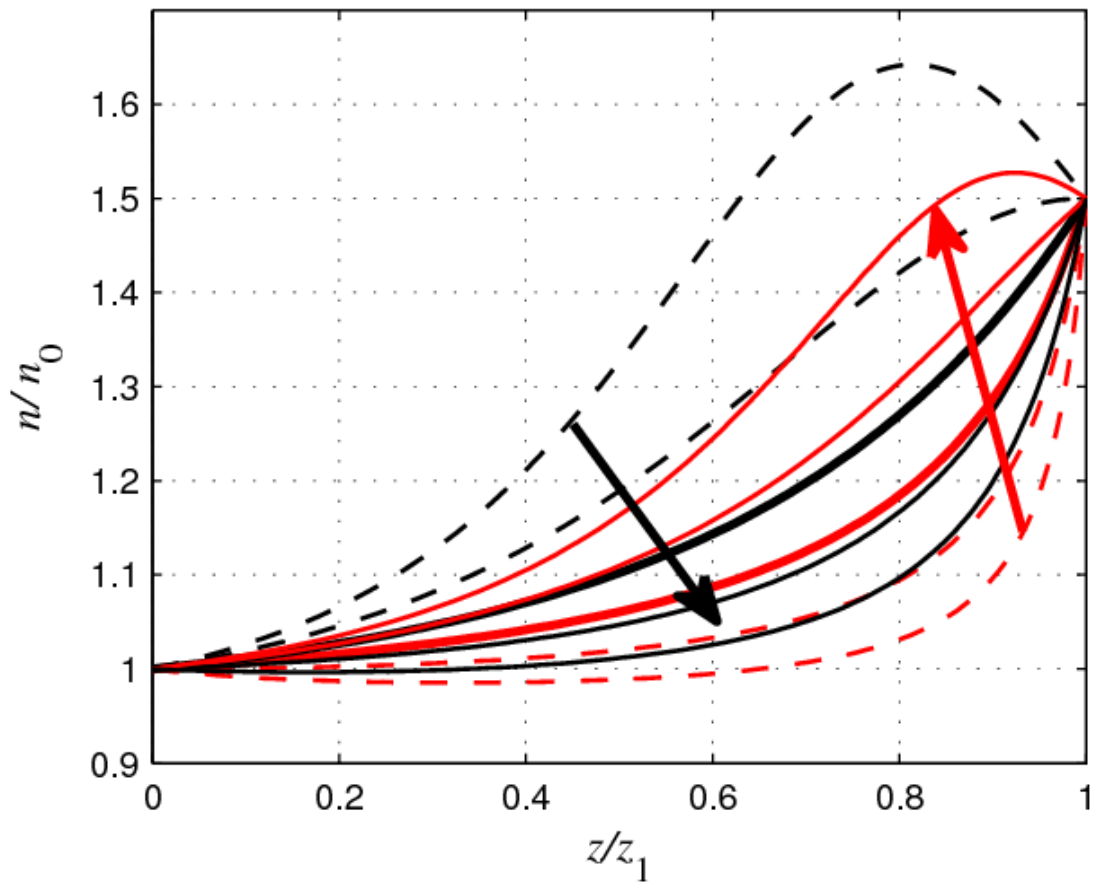

Figure 4. Same as in Figure 3 for an incidence of 70°, TE polarization and $n_a = n_0$.

Recall that the profiles with higher value of $|\beta\xi_1^2|$ present a higher curvature (see Figure 2, Figure 3 and Fig. 4). Figure 5 reveals that they indeed yield a higher reflectance (excluding some specific wavelengths where the reflectance drops to very low values). One important result is that, for any value of $\beta\xi_1^2$, the reflectance spectra of $\langle E \rangle$-form and $\langle H \rangle$-form profiles *perfectly overlap*. This equivalence was also verified analytically (not presented here). The profiles of $\langle E \rangle$-form and those of $\langle H \rangle$-form are thus *indistinguishable* from the reflectance (or transmittance) point of view, irrespective of the wavelength. Profiles presenting this peculiar behavior will be called *spectrally indistinguishable profiles* (SIPs). This result is particularly remarkable since, as revealed in Figure 2, profiles of a given pair of SIPs show significant differences: in particular, when one of them is concave, the other one is convex (except for very low values of $|\beta\xi_1^2|$). To our best knowledge these SIPs were not reported before.

Complementary simulations revealed that a stacking made of N identical $\langle E \rangle$-form layers and a stacking made of N identical $\langle H \rangle$-form layers (all sharing the same shape-parameter $\beta\xi_1^2$ value) also constitute a pair of SIPs. These findings illustrate very clearly, if proof were needed, that the inverse scattering problem, which consists in retrieving a refractive index profile from a reflectance spectrum, might not have a *unique* solution.

We have seen that, based on a *constant potential*, we can generate two families of solvable index profiles that are defined with *two or three* parameters. These profiles are either downward-convex or bell-shaped (the latter are roughly downward-concave; however one or two inflexion points may also be present). Hence, once the refractive index is specified at the right and left edges of a graded layer, there is one parameter left for adjusting the profile curvature. Convex and concave profiles of either $\langle E \rangle$-form and $\langle H \rangle$-form can be joined together like in [10] to get a composite periodic profile approaching a sinusoidal function. Indeed, by adjusting the third parameter of each convex and concave sub-profile one can manage to get the same slope on both sides of each junction nodes as illustrated in [10]. This adjustment however freezes the ratio of the positive to negative half-cycle amplitudes to a specific value. More flexibility is desired; this will be achieved with other solvable profiles as obtained after performing a Darboux transformation, as explained later.

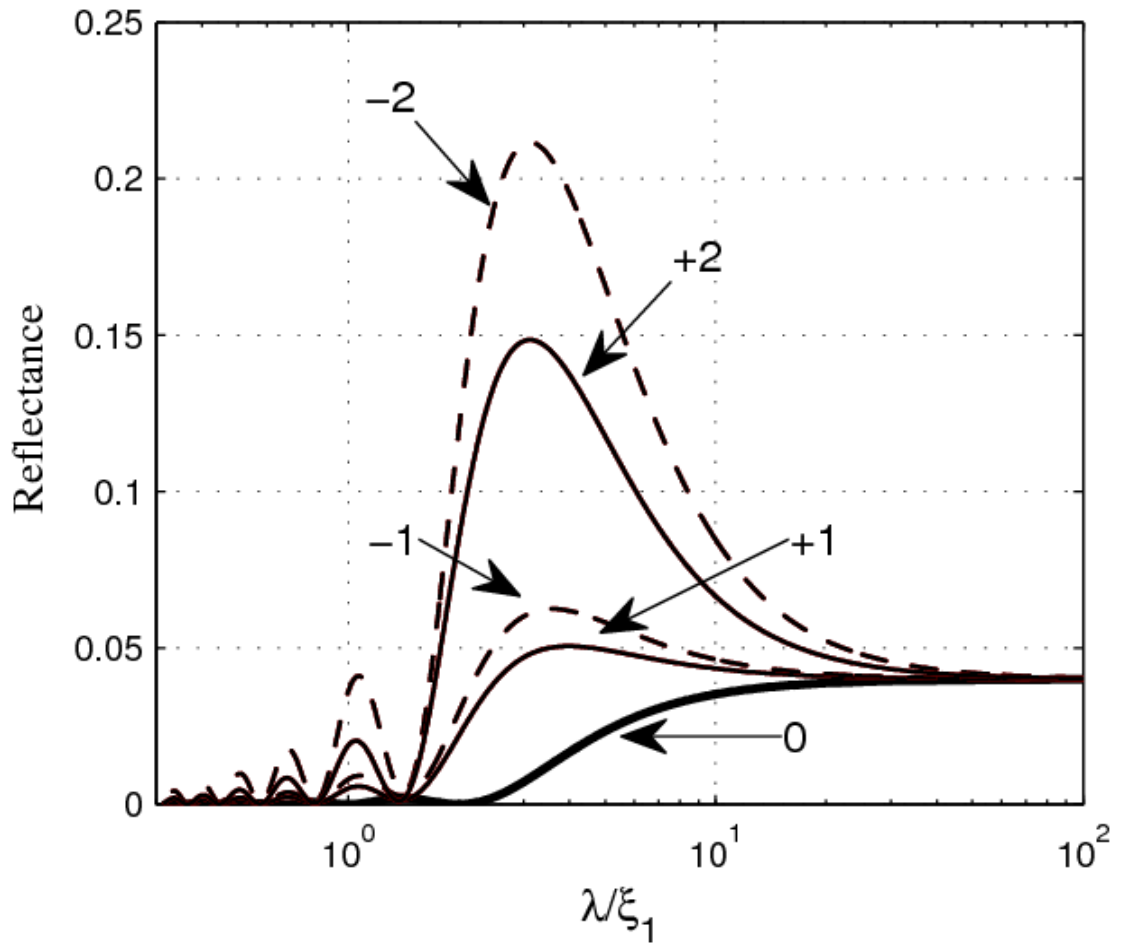

Figure 5. Reflectance spectra of the graded layers represented in Figure 2, Figure 3 and Fig. 4. $\lambda$ is the free-space wavelength and $\xi_1$ is the layer optical-thickness. The values of the shape-parameter $\beta\xi_1^2$ are indicated with arrows. There is index matching between the graded layer, the incidence medium and the substrate. For each value of $\beta\xi_1^2$, the reflectance spectra of the $\langle E \rangle$-form profile and of the $\langle H \rangle$-form profile overlap (resp. black and red profiles in Figure 2, Figure 3 and Fig. 4).

## 5. Applications of the PROFIDT method

### *5.1. One single PROFIDT with seed-potential* $V_{[0]} = 0$ *and transformation parameter* $p_1 = 0$

Starting with a zero potential (§4.1) and choosing a zero transformation parameter, $p_1 = 0$, means selecting one of the linear profiles of $s_{[0]}(\xi)$ in eq. (35) for the transformation function, i.e. $\tilde{\psi}_{[0,p_1=0]} \propto [1 \quad \xi]^t$. Its logarithmic derivative can be expressed as $\sigma_{[0,p_1=0]} = (\xi+\gamma)^{-1}$, where $\gamma$ is a free-parameter to be chosen in $]-\infty,-\xi_1[\cup]0,+\infty[$, where $\xi_1$ is the layer optical-thickness, for avoiding a singularity at $\xi = -\gamma$. According to eq. (A.4), the potential will be transformed by the DT into $V_{[1]} = 2(\xi+\gamma)^{-2}$.

The solution obtained after a first DT (see Annex A) is expressed as a LC of the independent solutions $K_{[1]}$, $P_{[1]}$ obtained from eq. (A.5):

$$\psi_{[1]} \propto \begin{bmatrix} K_{[1]} \\ P_{[1]} \end{bmatrix} \equiv \begin{bmatrix} K_{[0]}' - \sigma K_{[0]} \\ P_{[0]}' - \sigma P_{[0]} \end{bmatrix} = \begin{bmatrix} (ik_0 - \sigma)C^{(0)} \\ (-ik_0 - \sigma)S^{(0)} \end{bmatrix}, \tag{48}$$

where, for short, $\sigma \equiv \sigma_{[0,p_1=0]}$ and the initial independent functions $K_{[0]}$, $P_{[0]}$ correspond to the classical ones expressed in eq. (34). The associated admittance profiles are obtained by applying the transformation in eq. (A.8) since $p_1 = 0$:

$$s_{[1]}(\xi) = \eta_{[1]}^{\pm 1/2}(\xi) \propto \begin{bmatrix} B_{[1]} \\ D_{[1]} \end{bmatrix} \equiv \begin{bmatrix} (\xi+\gamma)^{-1} \\ (\xi+\gamma)^{2} \end{bmatrix}. \tag{49}$$

The $\langle E \rangle$-form profiles and the $\langle H \rangle$-form profiles are then obtained by substituting in eq. (28) the present definition of the LIS $B_{[1]}(\xi)$ and $D_{[1]}(\xi)$. The main results regarding the present solvable profiles are summarized in Table 1, case c2.

At the initial stage, the admittance profiles $\eta_{[0]}^{\pm 1/2}(\xi)$ were linear profiles in $\xi$, parameterized with *two* arbitrary constants (see eq. (35)), namely with the admittance values at the boundaries of the considered layer: $\eta_0$ at $\xi = 0$ and $\eta_1$ at $\xi = \xi_1$. After one PROFIDT, the new profiles $\eta_{[1]}^{\pm 1/2}(\xi)$ are rational functions in $\xi$ (see eq. (49)) that are parameterized with *three* arbitrary constants: $\eta_0$, $\eta_1$ and $\gamma$ (the latter one has a non-linear influence).

For illustrating the inverse-Liouville transform we again consider the "optical-TE" case. The $\langle E \rangle$-form profiles thus lead to an *explicit* expression of $n^*_{[1]}(z)$. First, one takes advantage of the fact that $B_{[1]}(\xi)^2 D_{[1]}(\xi) = 1$ for defining the required function $f_{BD}$ in eq. (C.11)-(C.12). After some manipulations, we get a degree-3 polynomial function of $z/z_1$ for $n^{*\,-3/2}_{[1]}(z)$:

$$n^*_{[1]}(z) = n^*_0(1+\kappa_1 z)^{-4/3}(1+\kappa_2 z)^{-2/3} \; ; \quad \begin{cases} \kappa_1 = \left( (n^*_1/n^*_0)^{-1/2}(\xi_1/\gamma+1)^{-1} - 1 \right) / z_1 \\ \kappa_2 = \left( (n^*_1/n^*_0)^{-1/2}(\xi_1/\gamma+1)^{2} - 1 \right) / z_1 \end{cases} \tag{50}$$

On the contrary, the $\langle H \rangle$-form profiles are defined in the $z$-space through an *implicit* relation between $\xi$ and $z$. From eq. (C.5)-(C.8) and the present definition of $B_{[1]}(\xi)$ and $D_{[1]}(\xi)$ we get:

$$z = -A_B^{\,2}\left[ (\xi+\gamma)^{-1} - \tau^{-1} \right] + \frac{A_D^{\,2}}{5}\left[ (\xi+\gamma)^5 - \gamma^5 \right] + A_B A_D \left[ (\xi+\gamma)^2 - \gamma^2 \right], \tag{51}$$

$$\begin{cases} A_B = \left( n_0^{*-1/2}(\xi_1+\gamma)^2 - n_1^{*-1/2}\gamma^2 \right) / \Delta \\ A_D = \left( -n_0^{*-1/2}(\xi_1+\gamma)^{-1} + n_1^{*-1/2}\gamma^{-1} \right) / \Delta \end{cases} \; ; \Delta = \left( \gamma^{-1}(\xi_1+\gamma)^2 - (\xi_1+\gamma)^{-1}\gamma^2 \right) \tag{52}$$

A few illustrative index profiles are presented in Figure 6 for the dielectric case while assuming normal incidence. As a reminder, the seed profiles are in bold. By playing with the $\gamma$ value, one can generate concave or convex profiles, much like by playing with the potential-value $\beta$ of the fundamental solutions (compare with Figure 3). The profiles in Figure 6 are nevertheless different from those in Figure 3. For both families, the index profiles are defined with three free-parameters.

A noticeable point is that the PROFIDT method gives rise to a new type of *spectrally indistinguishable profiles*. They are obtained with pairs of $\langle E \rangle$-form and $\langle H \rangle$-form profiles with same optical thickness $\xi_1$ and whose reduced parameters $\gamma_{(E)}/\xi_1$ and $\gamma_{(H)}/\xi_1$ satisfy the relation:

$\gamma_{(E)}/\xi_1 + \gamma_{(H)}/\xi_1 = -1$ (in addition to an index matching condition at both sides of each graded layer). When plotted vs. the reduced wavelength $\lambda/\xi_1$, the spectra indeed overlap, see Figure 7. On the other side, one can notice a slight difference between the reflectance spectra of $\langle E \rangle$-form and $\langle H \rangle$-form profiles that share *a common* $\gamma/\xi_1$ value. This difference vanishes when the index is set equal at both ends of each layer, i.e. $n_1 = n_0$. In this situation, all *four* profiles of $\langle E \rangle$ and $\langle H \rangle$-form whose reduced parameter is taken from a pair of $\gamma/\xi_1$-values that sum up to -1 are *spectrally indistinguishable.*

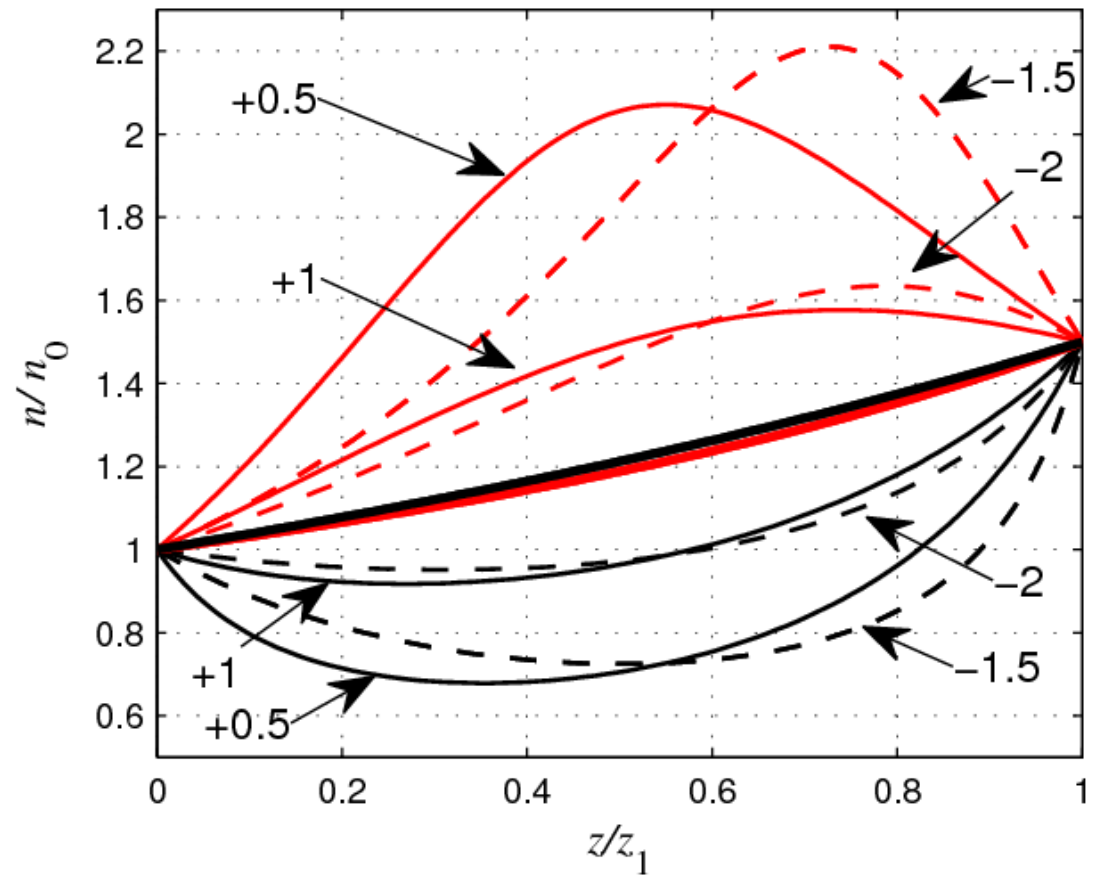

Figure 6. Solvable index profiles obtained after one PROFIDT, starting from zero-potential and choosing a zero transformation parameter. Particular case with $n_1/n_0 = 1.5$ and normal incidence. The thin line curves correspond to different values of the third free-parameter $\gamma$ (plain curves correspond to positive values of $\gamma$, dashed curves to negative values; the values of the normalized parameter $\gamma/\xi_1$ are shown in the figure, where $\xi_1$ is the optical thickness of the layer). The bold line curves correspond to the seed solutions, i.e. the fundamental solutions with zero-potential. In black : $\langle E \rangle$-form profiles, in red : $\langle H \rangle$-form profiles.

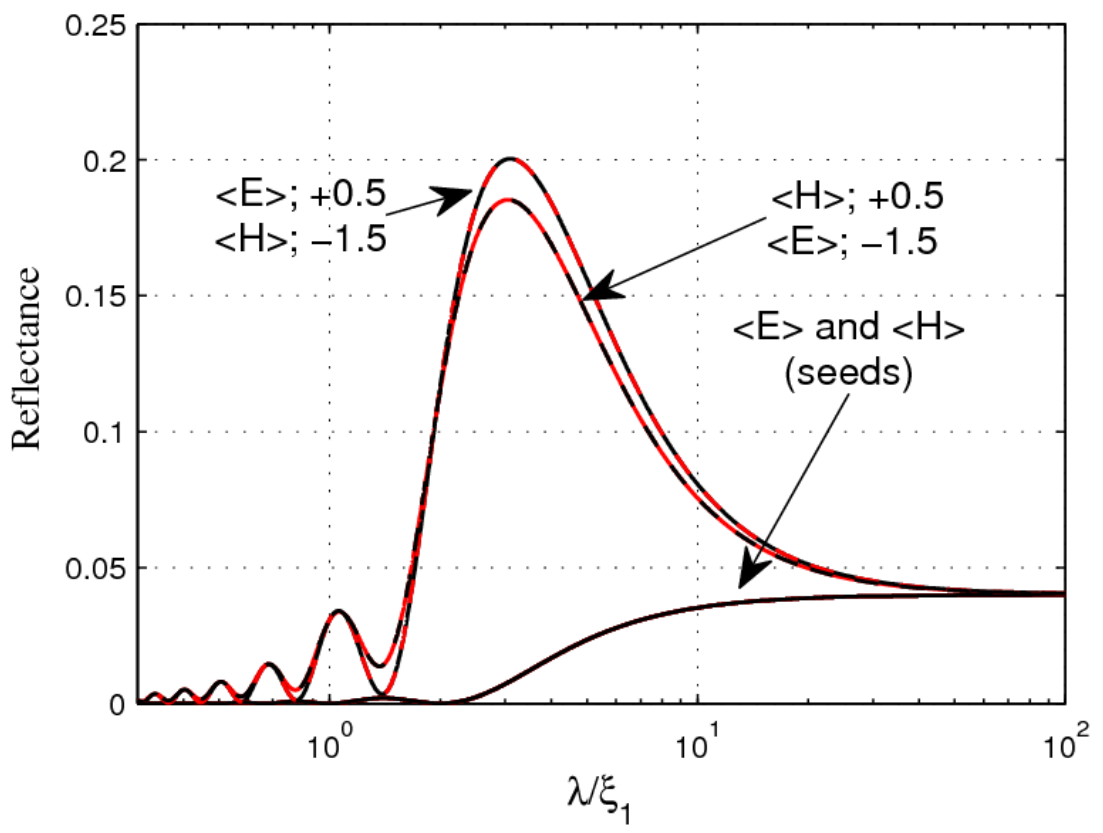

Figure 7. Reflectance spectra of the profiles in Figure 6 with $\gamma/\xi_1$ =0.5 or -1.5 as indicated in the figure. As a reminder, we also reported the (overlapping) spectra of the $\langle E\rangle$-form and $\langle H\rangle$-form seed solutions (zero-potential). The front side refractive-index is equal to the incidence index ($n_0 = n_a$) and the back side index is equal to the substrate index ($n_1 = n_s$). Pairs of $\langle E\rangle$-form (in black) and $\langle H\rangle$-form (in red) reflectance spectra overlap when the reduced parameters $\gamma_{(E)}/\xi_1$ and $\gamma_{(H)}/\xi_1$ sum up to -1.

### *5.2. One single PROFIDT with seed-potential* $V_{[0]} = 0$ *and transformation parameter* $p_1 \neq 0$

The admittance profiles and the wave functions obtained from a null seed-potential $V_{[0]}$=0 and a non-null transformation parameter are summarized in Table 1, cases c3 and c4 for $p_1$>0 as well as case c5 for $p_1$<0. The new potential $V_{[1]}(\xi)$ has three different forms which belong to the Darboux-Pöschl-Teller potential family; the first one is also called the one-soliton potential. This yields six classes of solvable admittance profiles. Although these profiles now involve *four* free-parameters, i.e. one more as compared to the fundamental profiles in Fig. 2, their shapes prove to be globally similar, i.e. either concave or convex. The addition of a fourth parameter didn't bring much variety. For this reason, profiles that are suitable for modelling more complex finite-thickness layers will be described in §5.4.

However, when applied to the *half-space* case, the present class of profiles shows interesting properties. For the sake of boundedness we have to restrict to the case c3 ($p_1$>0) and to the first independent solution of $s(\xi)$, which means:

$$s_{[1]}(\xi) = \eta_{[1]}{}^{\pm 1/2}(\xi) \propto \tanh(\hat{\xi}) \quad ; \quad \hat{\xi} = \xi/\xi_c + \gamma \tag{53}$$

where $\xi_c$ is the characteristic optical-thickness of the graded semi-infinite medium (it is itself defined from the transformation parameter $p_1$ by $\xi_c \equiv p_1{}^{-1/2}$). Furthermore, $\gamma$ is another free-parameter which can be related to the admittance value $\eta_0$ at the surface $\xi = 0$ and to the asymptotic value $\eta_\infty$ "deep" inside the material (i.e. for $\xi \to \infty$) by $\gamma \equiv \mathrm{atanh}(\eta_0/\eta_\infty)^{\pm 1/2}$. Actually, the solvable profiles in eq. (53) of both $\langle E \rangle$-form and $\langle H \rangle$-form merge into a single expression where the $j = \pm 1$ parameter value determines whether the profile is increasing or decreasing:

$$\eta_{[1]}{}^{1/2}(\xi) = \eta_\infty{}^{1/2} \tanh^j\left(\xi/\xi_c + \mathrm{atanh}(\eta_0/\eta_\infty)^{j/2}\right), \quad j = \begin{cases} +1 & \text{if} \quad \eta_0 \le \eta_\infty \\ -1 & \text{if} \quad \eta_0 > \eta_\infty. \end{cases} \tag{54}$$

These profiles will be called of $\tanh^{\pm 1}(\hat{\xi})$-type. Solvable profiles of this type were already found in [52] in the course of solving variable-coefficient PDE by Bäcklund transformations (namely when solving the wave equation in the time domain). Truncated $\tanh^{\pm 1}(\hat{\xi})$-type profiles were obtained in [11] in the context of acoustic waves (3rd and 4th cases). We will now develop a little more about their representation in the physical coordinate space, about the related surface admittance and about a new solvable profile obtained by stitching together two $\tanh^{\pm 1}(\hat{\xi})$-type profiles.

In the "optical-TE" case, after defining a characteristic depth $z_c$ according to $z_c \equiv \xi_c / n_0^*$, the pseudoindex profiles are described through a *single implicit* expression involving the parameter $\xi$, should $n_\infty^*$ be higher or lower than $n_0^*$:

$$\frac{z}{z_c} = \frac{\xi}{\xi_c}\frac{n_0^*}{n_\infty^*} - \frac{n_0^*}{\sqrt{n_\infty^* n_{[1]}^*(\xi)}} + \sqrt{\frac{n_0^*}{n_\infty^*}}. \tag{55}$$

Eq. (54) is actually invertible: the parameter $\xi$ obtained therefrom can thus be substituted into eq. (55) to get an *implicit* closed-form expression that defines the pseudo-index profile vs. position $z$ :

$$\frac{z(n_{[1]})}{z_c} = \frac{n_0^*}{n_\infty^*}\left(\operatorname{atanh}\left(\frac{n_{[1]}^*}{n_\infty^*}\right)^{j/2} - \operatorname{atanh}\left(\frac{n_0^*}{n_\infty^*}\right)^{j/2}\right) - \frac{n_0^*}{\sqrt{n_\infty^* n_{[1]}^*}} + \sqrt{\frac{n_0^*}{n_\infty^*}} \ , \quad j = \begin{cases} +1 & n_0 \le n_{[1]} < n_\infty \\ -1 & n_0 \ge n_{[1]} > n_\infty \ . \end{cases} \tag{56}$$

A series of solvable profiles of $\tanh^{\pm 1}(\hat{\xi})$-type is reported in Figure 8 vs. the normalized optical depth (or vs. the normalized physical depth, after an inverse-Liouville transformation) for different values of the index ratio $n_\infty^* / n_0^*$ ranging from 0.5 to 2.

For evaluating the EM-field solution, we first use the fact that for $z \to +\infty$ there should be only a forward propagating wave component for $E$ and $H$ , which, after a multiplication by $\eta_{[1]}^{\pm 1/2}$ , should also apply to $\psi_{[1]}$ since $\eta_{[1]} \to \eta_\infty$ . The solution for $\psi_{[1]}$ is thus reduced to a single term:

$$\psi_{[1]} \propto (ik_0 + \sigma) S^{(0)} . \tag{57}$$

When substituting $\psi_{[1]}$ and $s_{[1]}$ into eq. (23)-(24), we obtain an extremely simple expression for the surface admittance:

$$Y = \left(\frac{H}{E}\right)_{\xi=0} = \eta_0 \frac{ik_0 \xi_c + x_0^{-1}}{ik_0 \xi_c + x_0} . \tag{58}$$

where $x_0 = \sqrt{\eta_0 / \eta_\infty}$ . As expected, the asymptotic value of the surface admittance for an increasing frequency is $\eta_0$ i.e. the surface tilted-admittance, whereas for a decreasing frequency, the asymptotic value is $\eta_\infty$ i.e. the bulk tilted-admittance. When the admittance is continuous at the interface with the incidence material, i.e. $\eta_a = \eta_0$ , the reflectance of the graded layer reduces to:

$$R = \frac{\left(x_0 - x_0^{-1}\right)^2}{\left(x_0 + x_0^{-1}\right)^2 + 4\left(k_0 \xi_c\right)^2}. \quad (59)$$

The Fresnel reflectance value relative to the admittance step $\eta_0 \to \eta_\infty$ is well recovered for asymptotically large values of the wavelength. On the other side, for small wavelengths, the reflectance evolves like $\lambda^2$. This is in accordance to the general trend reported in [46], [53], [54]: in presence of a $C^{m-1}$ transition profile (i.e. when the (m-1)$^{th}$ derivative is continuous but not the m$^{th}$), the reflectance spectrum is expected to evolve like $\lambda^{2m}$ for small values of the wavelength. Since the profiles are simply continuous, not derivable, at the free surface, the expected (and indeed observed) trend is actually $\lambda^2$. In the end, the $\tanh^{\pm 1}\left(\hat{\xi}\right)$-type profiles become "reflectionless" for a vanishing wavelength. The other interesting point is that replacing $x_0$ in eq. (59) by its inverse does not modify the reflectance value. This means that two graded media of $\tanh^{\pm 1}\left(\hat{\xi}\right)$-type whose admittance values $\eta_\infty / \eta_0$ are reciprocal one to the other are *spectrally indistinguishable*: the reflectance spectra, when normalized by the corresponding $\xi_c$ value, are strictly equivalent (yet, the admittance profiles are profoundly different since one is increasing whereas the other one is decreasing). All these features are observed in Figure 9 which reports the reflectance spectra at normal incidence for the $\tanh^{\pm 1}\left(\hat{\xi}\right)$-type semi-infinite profiles displayed in Figure 8.

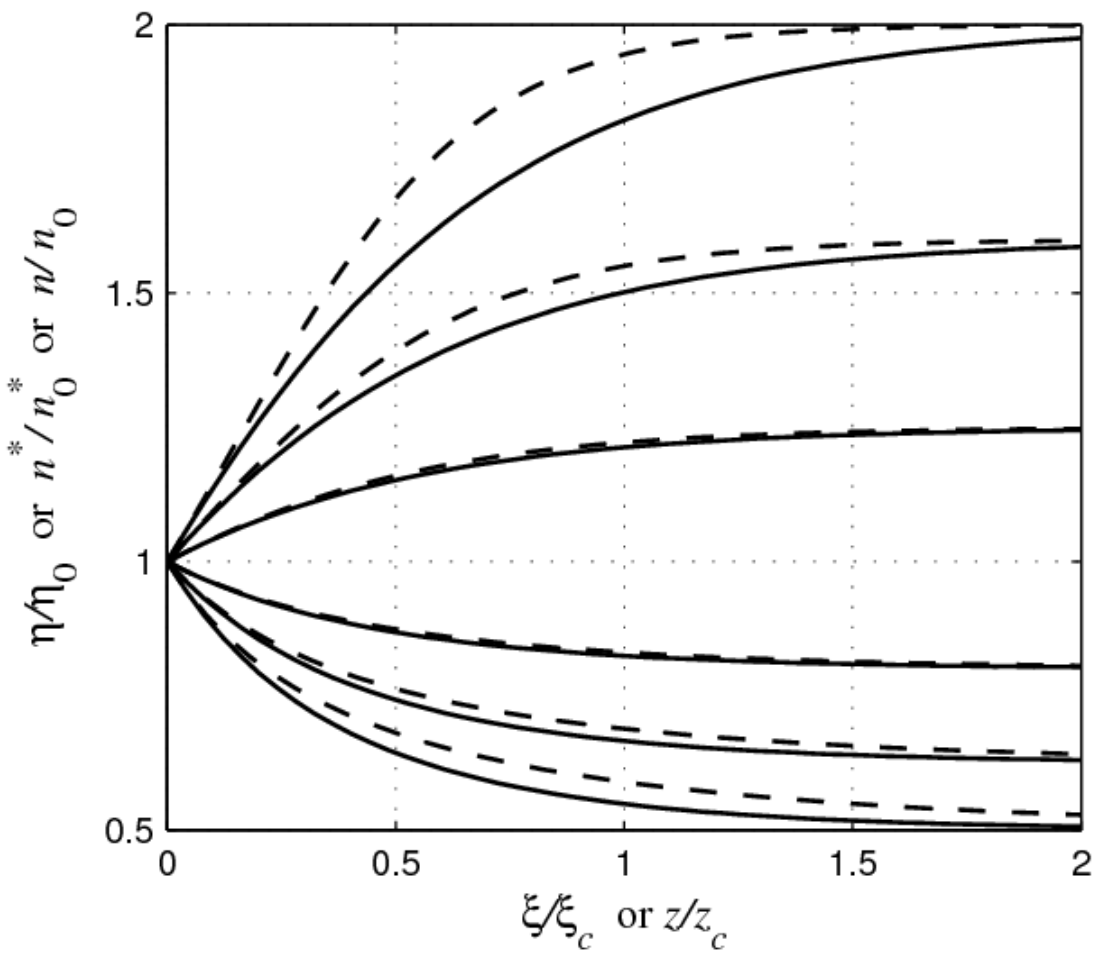

Figure 8. Solvable profiles of $\tanh^{\pm 1}(\hat{\xi})$-type for different values of the bulk-to-surface index ratio (from bottom to top: $\eta_\infty/\eta_0$ = 0.5, 0.625, 0.8, 1.25, 1.6, 2. The solid line curves represent the tilted admittance $\eta(\xi)/\eta_0$ or the tilted index $n^*(\xi)/n_0^*$ vs. the normalized optical-depth $\xi/\xi_c$ whereas the dashed curves represent, after an inverse-Liouville transformation (while assuming normal incidence), the normalized index $n/n_0$ vs. the normalized physical-depth $z/z_c$ . $\xi_c$ (resp. $z_c$ ), is the characteristic optical thickness (resp. the characteristic physical thickness) of the corresponding graded profile.

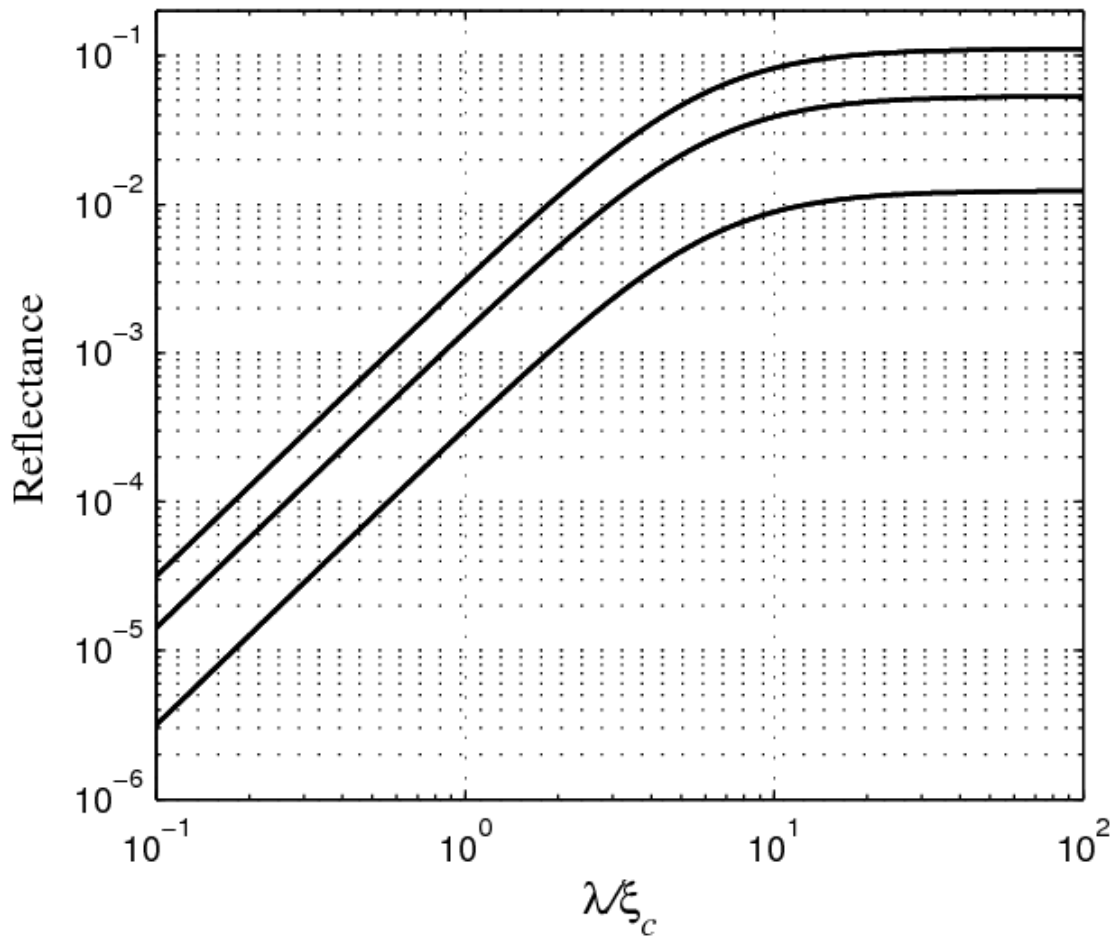

Figure 9. Reflectance spectra of the semi-infinite graded layers of $\tanh^{\pm 1}(\hat{\xi})$-type presented in Figure 8 when joined to an incident medium with $n_a = n_0$. The characteristic optical thickness $\xi_c$ is used for normalizing the free-space wavelength $\lambda$ .The spectra of the profiles corresponding to $n_\infty/n_0$ = 2 and $n_\infty/n_0$ =0.5 are overlapping (uppermost curve). Same for the profiles with $n_\infty/n_0$ = 1.6 and 0.625 (middle curve), resp. with $n_\infty/n_0$ = 1.25 and 0.8 (bottom curve).

Let us mention that the potential-function described in Table 1, case c3, when expressed in the physical-coordinate space and spreading over the *whole real line*, belongs to the family of *reflectionless* potentials [55].

The $\tanh^{\pm 1}(\hat{\xi})$-type profiles are defined with merely *three* parameters, which means that once the surface and bulk properties are assigned, there is only one parameter left for additional adjustments. The transition depth $z_c$ and the index slope at the surface are indeed correlated. For breaking this limitation, one could repeat the PROFIDT procedure and hence get more “flexible” solutions, namely with more adjustable parameters; this will be shown in the next paragraph.

One can play further with the $\tanh^{\pm 1}(\hat{\xi})$-type profiles by joining two of them, one over the positive half-line the other one over the negative half-line. The respective parameters can be adjusted so that the composite profile is continuous at the junction node up to the first derivative. For this purpose, the two half-space profiles should be antisymmetric with respect to the interface, i.e. one of them should be of tanh-type whereas the other one should be of coth-type. The composite profile is then defined according to (the interface is shifted to the position $\xi = \xi_i$):

$$\eta^{1/2}(\xi) = \begin{cases} \eta_{+\infty}^{\ 1/2} \tanh^{j}\left((\xi - \xi_i)/\xi_{c+} + \operatorname{atanh} x_{i+}^{\ j}\right); & x_{i+} = (\eta_i/\eta_{+\infty})^{1/2}; \quad \xi \geq \xi_i \\ \eta_{-\infty}^{\ 1/2} \tanh^{-j}\left((\xi_i - \xi)/\xi_{c-} + \operatorname{atanh} x_{i-}^{\ -j}\right); & x_{i-} = (\eta_i/\eta_{-\infty})^{1/2}; \quad \xi < \xi_i \end{cases} \tag{60}$$

The parameter $j$ is set to +1 or -1 for getting a monotonous increasing (resp. decreasing), profile over the entire line (other combinations, i.e. with the same exponent left and right of the interface, lead to profiles with either a top-oriented corner or a bottom-oriented corner at the interface; we will not consider them further).

For a profile as described in eq. (60), the slope continuity is achieved provided the characteristic optical-thicknesses of the left and right parts, $\xi_{c-}$ ( resp. $\xi_{c+}$), satisfy the following relation:

$$\xi_{c-}\left(x_{i+}^{\ -1} - x_{i+}\right) = \xi_{c+}\left(x_{i-} - x_{i-}^{\ -1}\right) \tag{61}$$

Notice that in the case of an equal logarithmic rise of the admittance, left and right of the interface (i.e. $x_{i+} = x_{i-}^{\ -1}$) the previous relation becomes simply $\xi_{c-} = \xi_{c+}$. The composite profiles expressed in eq. (60), when truncated at $\xi = 0$, describe fairly well the case of a coated semi-infinite substrate with a diffuse interface. This structure is fully described by providing the bulk admittance $\eta_{+\infty}$, the admittance at the interface $\eta_i$, the surface admittance $\eta_0$ and the right optical-thickness $\xi_{c+}$ which characterizes the interface diffusion extent. The (virtual) admittance at $\xi \to -\infty$, i.e. $\eta_{-\infty}$ is determined from the former data by expressing eq. (60) at the free surface, i.e. $\eta_0^{\ 1/2} = \eta_{-\infty}^{\ 1/2} \tanh^{-j}\left(\xi_i/\xi_{c-} + \operatorname{atanh} x_{i-}^{\ -j}\right)$, which implies to solve a transcendental equation while also taking into account eq. (61). One could instead specify the (virtual) admittance $\eta_{-\infty}$ beforehand; the surface admittance is then inferred explicitly from the former relation. Figure 10 shows joint coth/tanh profiles (hence rising profiles) obtained by imposing the constraint of a continuous slope at the interface and by varying the thickness of the interface diffusion layer. Similarly, decreasing synthetic profiles may be obtained by joining tanh and coth profiles instead.

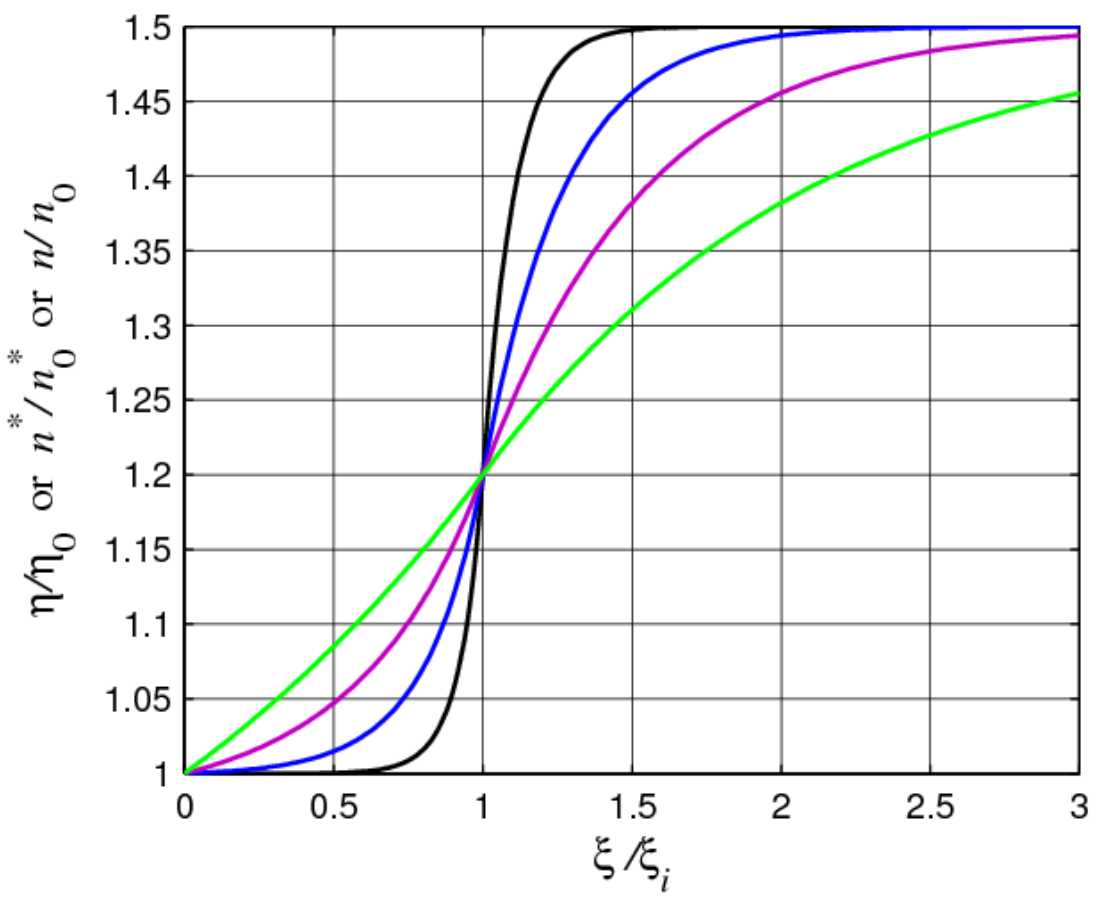

Figure 10. Bilayer-profiles made of a coating of coth-type and optical thickness $\xi_i$ laid over a semi-infinite profile of tanh-type. The characteristic optical thicknesses are resp. $\xi_{c-}$ and $\xi_{c+}$. They are related through eq. (61) and they determine the optical thickness of the interface "diffusion layer". A progressively thicker interface layer is obtained when setting $\xi_{c+}/\xi_i$ to 0.2 (black), 0.5 (blue), 1 (magenta) and then 2 (green).

The optical response of the bilayer profile is obtained by applying the recursive surface admittance relation in eq. (B.5), starting from the semi-infinite layer admittance expressed in eq. (58) and substituting in eq. (B.5) the four matrix coefficients related to the (truncated) front layer. These coefficients are obtained by substituting in eq. (B.1)-(B.2) the expression of the linearly independent solutions $K(\xi, k_0)$, $P(\xi, k_0)$ in Table 1, case c3 (tanh-type) or case c4 (coth-type). After tedious algebraic manipulations (which have been alleviated and validated with the use of a symbolic-computation tool) we get the following compact expressions:

$$\begin{bmatrix} A \\ B\left(im\eta_{-\infty}{}^{-1}\right)^{-1} \\ C\left(im\eta_{-\infty}\right)^{-1} \\ D \end{bmatrix} = \frac{1}{\tilde{k}_0{}^2+1}\begin{bmatrix} \tilde{k}_0{}^2 x_{0-}{}^{-1}x_{i-}+1 & \tilde{k}_0\left(x_{0-}{}^{-1}-x_{i-}\right) \\ \tilde{k}_0\left(x_{i-}{}^{-1}-x_{0-}{}^{-1}\right) & \tilde{k}_0{}^2 x_{0-}{}^{-1}x_{i-}{}^{-1}+1 \\ \tilde{k}_0\left(x_{i-}-x_{0-}\right) & \tilde{k}_0{}^2 x_{0-}x_{i-}+1 \\ \tilde{k}_0{}^2 x_{0-}x_{i-}{}^{-1}+1 & \tilde{k}_0\left(x_{0-}-x_{i-}{}^{-1}\right) \end{bmatrix} \times \begin{bmatrix} \cos\left(k_0\xi_i\right) \\ \sin\left(k_0\xi_i\right) \end{bmatrix}, \tag{62}$$

with $x_{0-} = \left(\eta_0/\eta_{-\infty}\right)^{1/2}$ and $\tilde{k}_0 = k_0\xi_{c-}$.

The reflectance spectra of the bilayers in Figure 10 are reported in Figure 11, assuming index matching at the free surface. For large values of the wavelength we again get the Fresnel reflectance value. The reflectance again drops for a vanishing wavelength. The logarithmic slope however depends on the thickness of the "diffusion layer". For a rather thick diffusion layer (e.g. $\xi_{c+}/\xi_i$ =2 in Figure 10) there is essentially a *discontinuity of the slope* at the front surface; the consequence, for the reflectance, is that $R \propto \lambda^2$. For a thin diffusion layer (i.e. $\xi_{c+}/\xi_i \leq 0.5$), the essential discontinuity is a *discontinuity of the curvature* at the interface; the consequence is that $R \propto \lambda^4$. Furthermore, the reflectance is conditioned by $\xi_{c-}$ (and $\xi_{c+}$), not by $\xi_i$, which explains that the spectra in Figure 11 are then simply shifted towards the left. For a diffusion layer of the order of the coating optical thickness, the spectrum is much "noisy" and the logarithmic slope is intermediate.

Exactly the same reflectance spectra were obtained with *decreasing* profiles as generated by taking the reciprocal values of the dimensionless tilted-index reported in Figure 10 (i.e. with joint tanh/coth profiles). This result is in accordance with the previous observation that *'twin'* tanh and coth profiles (i.e. of same characteristic optical-thickness but reciprocal index-contrast) are *spectrally indistinguishable* (refer to Figure 9).

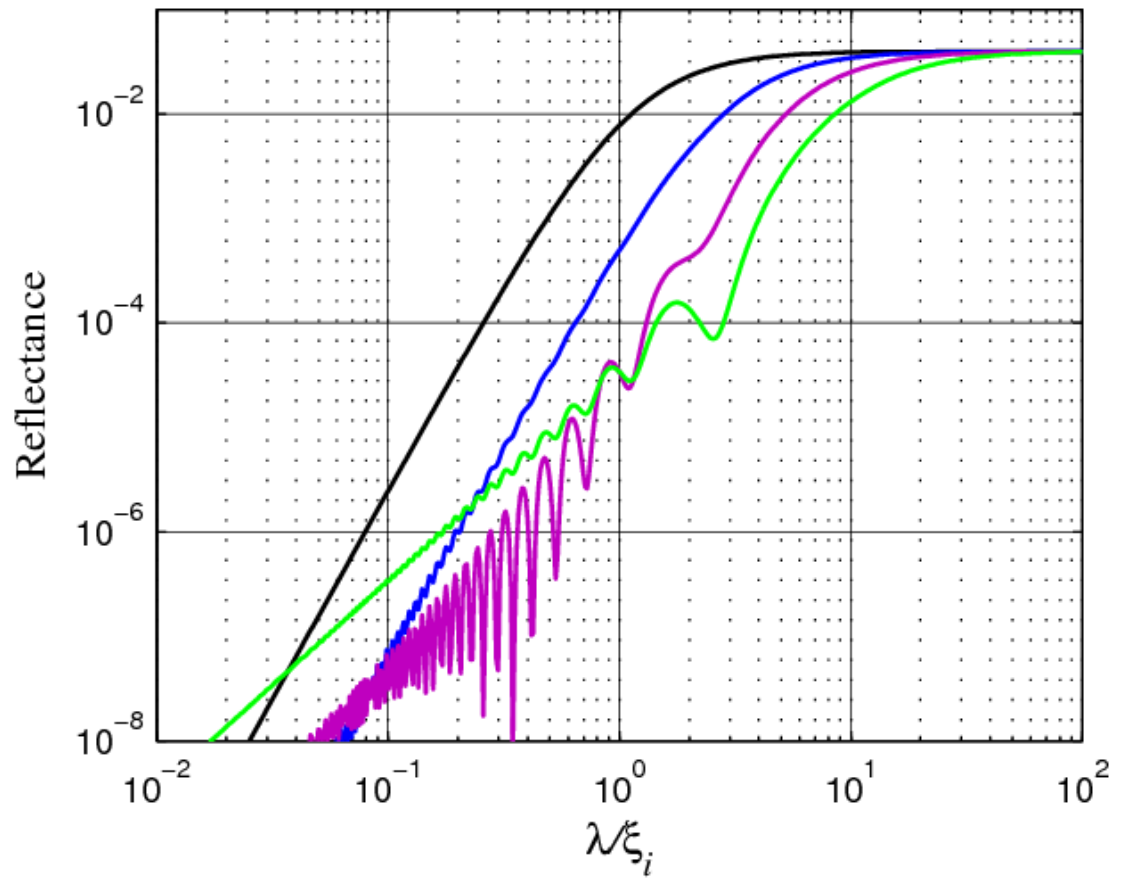

Figure 11. Reflectance spectra of the synthetic bilayer-profiles in Figure 10. The free-space wavelength $\lambda$ is divided by the coating optical-thickness $\xi_i$. The same spectra are obtained for the profiles that are reciprocal of those depicted in Figure 10.

The bilayer model, in its most general form, is a 9-parameter model, including the two $\pm 1$ exponents. The number of parameters reduces to 8 and 6 when the admittance is assumed continuous, resp. $C^1$ (i.e. differentiable to 1st order) at the interface. A 4-parameter $C^\infty$ sigmoidal model was described in [56] and the reflectance of the entire profile (i.e. over the full-line) has a particularly simple analytical expression. However, if that profile was truncated for the purpose of modelling a coated substrate like in Figure 10, the front surface reflectance would most probably feature hypergeometric functions since the related field-solution does so. This is in contrast to the simplicity of the reflectance expression as obtained from eq. (62) and eq. (58).

### 5.3. *Two-fold PROFIDT with seed-potential* $V_{[0]} = 0$ *and transformation parameters* $p_1 \neq p_2 \neq 0$

The solvable admittance profiles as obtained after a two-fold PROFIDT, starting with the seed-potential $V_{[0]} = 0$, can be expressed by [36]:

$$s_{[2]}(\xi) = \eta_{[2]}^{\pm 1/2}(\xi) \propto \begin{bmatrix} B_{[2]} \\ D_{[2]} \end{bmatrix} \equiv \begin{bmatrix} 0 \\ 1 \end{bmatrix} \frac{p_2 - p_1}{\sigma_{[0,p_2]} - \sigma_{[0,p_1]}} + \begin{bmatrix} 1 \\ \xi \end{bmatrix} \frac{p_1 \sigma_{[0,p_2]} - p_2 \sigma_{[0,p_1]}}{\sigma_{[0,p_2]} - \sigma_{[0,p_1]}}, \tag{63}$$

where each of the log-derivatives $\sigma_{[0,p_1]}$ and $\sigma_{[0,p_2]}$ may take, independently from the other, one among the three forms described in cases 3, 4 and 5 of Table 1. Each log-derivative $\sigma_{[0,p_k]}$ involves two parameters: $\xi_{c_k} \equiv |p_k|^{-1/2}$ and $\gamma_k$, from which are defined two transformed-optical-thicknesses: $\hat{\xi}_k = \xi/\xi_{c_k} + \gamma_k$, $k = 1,2$.

Let us now restrict to the semi-infinite layer case as we did in § 5.2 when establishing eq. (54). For this purpose, $p_1$ and $p_2$ must be positive and the log-derivatives $\sigma_{[0,p_k]}$ should restrict to the first two forms described as case 3 and case 4 in Table 1. As such, the new admittance takes the following expression:

$$\eta_{[2]}^{\pm 1/2}(\xi) = \eta_\infty^{\pm 1/2} \frac{\xi_{c2}/\xi_{c1} \tanh^{j_2}(\hat{\xi}_2) - \tanh^{j_1}(\hat{\xi}_1)}{\xi_{c2}/\xi_{c1} \tanh^{j_1}(\hat{\xi}_1) - \tanh^{j_2}(\hat{\xi}_2)}, \tag{64}$$

where the exponents $j_1$ and $j_2$ can take the values +1 or -1 independently, which implies four different combinations. The new admittance profile involves *five*-parameters: $\xi_{c_1}$, $\gamma_1$, $\xi_{c_2}$, $\gamma_2$ and $\eta_\infty$. Alternatively, expressing that, at the surface, $\eta_{[2]}(\xi = 0) = \eta_0$, yields the following relation between the parameters:

$$\xi_{c2}/\xi_{c1} = \frac{(\eta_\infty/\eta_0)^{\pm 1/2} \tanh^{j_1}(\gamma_1) - \tanh^{j_2}(\gamma_2)}{(\eta_\infty/\eta_0)^{\pm 1/2} \tanh^{j_2}(\gamma_2) - \tanh^{j_1}(\gamma_1)}, \tag{65}$$

which allows changing the set of *five* free-parameters: $\xi_{c_1}$, $\gamma_1$, $\xi_{c_2}$, $\gamma_2$ and $\eta_\infty$ for the set: $\xi_{c_1}$, $\gamma_1$, $\gamma_2$, $\eta_0$, and $\eta_\infty$. Hence, after specifying the admittance boundary values $\eta_0$, and $\eta_\infty$ together with the characteristic optical-thickness $\xi_{c1}$, there are two parameters left, namely $\gamma_1$ and $\gamma_2$, for adjusting at will the admittance profile shape (the remaining coefficient $\xi_{c2}$ is finally inferred from eq. (65)). Notice that for each parameter-set selection, several profiles can be built depending on the choice for the exponent signs in eq. (64)-(65): first, the $\langle E \rangle$-form or the $\langle H \rangle$-form is selected which determines the common sign of the exponent in $\eta_{[2]}^{\pm 1/2}$ and in $\eta_{\infty}^{\pm 1/2}$, secondly, there are four possibilities for the tanh/coth combinations as mentioned before. However, a permutation of the $\langle E \rangle$-form and $\langle H \rangle$-form profiles is observed when switching from the tanh/tanh to the coth/coth combination, similarly when switching from the tanh/coth to the coth/tanh combination. This means that despite four different $\langle E \rangle$-form profiles and four different $\langle H \rangle$-form profiles are obtained, there are altogether only *four* different curve-sets.

Figure 12 provides an insight to the diversity of profile shapes to which the formula in eq. (64) gives access. For this illustration, we have set the bulk-to-surface admittance ratio to $\eta_{\infty}/\eta_0$ =1.6. With this particular value, we got, after the first PROFIDT, a single profile, namely the second plain curve from top in Figure 8. After a second PROFIDT we now get plenty of curves with different shapes and initial slopes (even with negative slopes) and with possibly an overshoot or an undershoot. In Figure 12 we merely reported the $\langle E \rangle$-form and $\langle H \rangle$-form profiles obtained with the tanh/tanh combination (i.e. $j_1 = j_2$ =1).

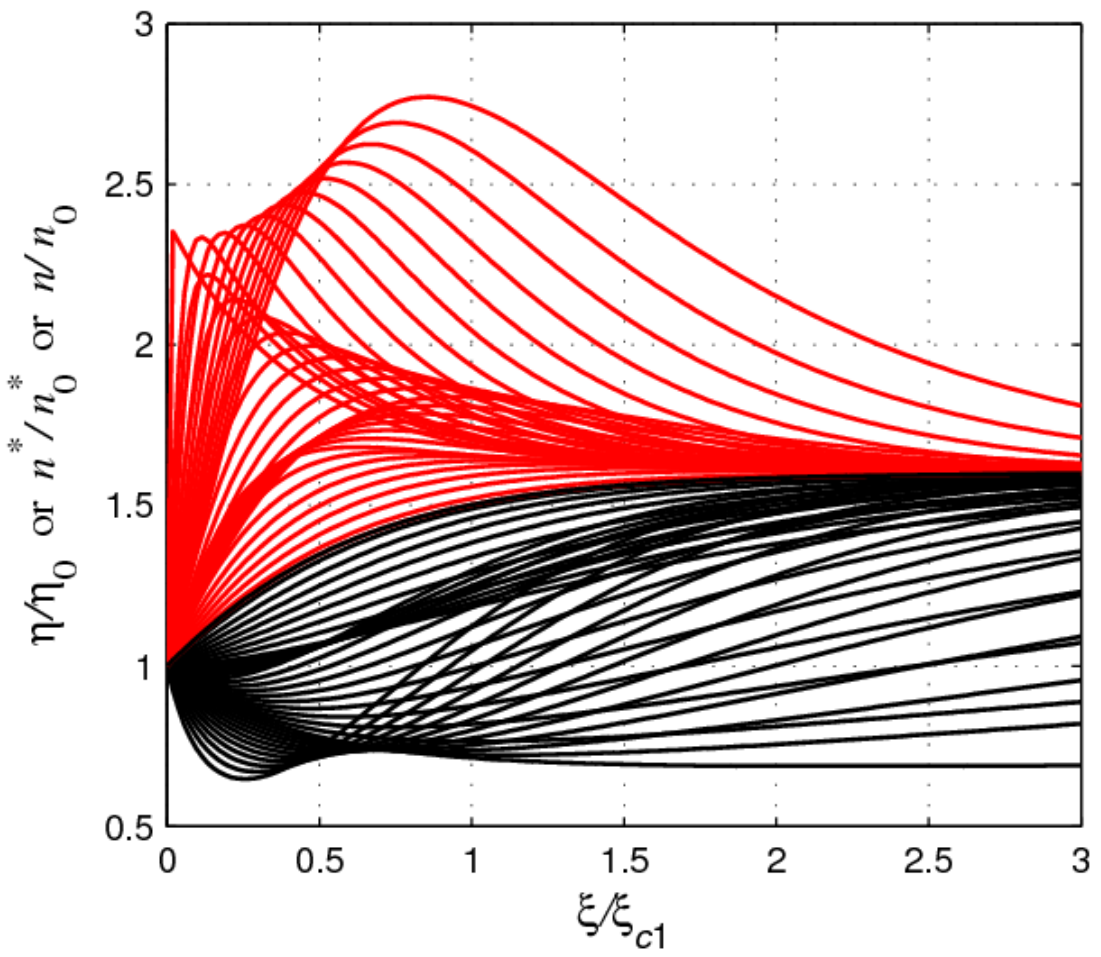

Figure 12. A selection of solvable index profiles obtained after a two-fold PROFIDT, when starting from a zero seed-potential. $\langle E \rangle$-form profiles are in black; $\langle H \rangle$-form profiles are in red (tanh/tanh combination). Particular case with $\eta_{\infty}/\eta_0$ =1.6. The frontier between the black and red curve-sets corresponds to the second (plain) curve from top in Figure 8 as obtained after one single PROFIDT. Any curve describes either the tilted admittance $\eta(\xi)/\eta_0$ (valid for both TE and TM modes and for any incidence) or the relative tilted index $n^*(\xi)/n_0^*$ (in the case of non-magnetic materials, $\mu_r(\xi)$=1) or the relative refractive index $n(\xi)/n_0$ (when further assuming normal incidence).

After some algebra we find the following expression for the surface admittance:

$$Y = \eta_0 \left( -k_0^2 \xi_{c1} \xi_{c2} + ik_0 \frac{\xi_{c1}{}^2 - \xi_{c2}{}^2}{\xi_{c1} T_2^{-1} - \xi_{c2} T_1^{-1}} + \sqrt{\frac{\eta_\infty}{\eta_0}} \right) \left( -k_0^2 \xi_{c1} \xi_{c2} + ik_0 \frac{\xi_{c1}{}^2 - \xi_{c2}{}^2}{\xi_{c1} T_2 - \xi_{c2} T_1} + \sqrt{\frac{\eta_0}{\eta_\infty}} \right)^{-1}, \qquad (66)$$

where the symbol $T_k$ is used for $\tanh^{j_k}(\gamma_k)$. We find again, as expected, that the asymptotic limit for increasing frequency, (resp. decreasing frequency), is the *surface* tilted-admittance $\eta_0$, resp. the *bulk* tilted-admittance $\eta_\infty$. The simplicity of this expression is particularly remarkable given the wide variety of shapes achieved with the profiles of order 2, as partially illustrated in Figure 12.

A four-parameter solvable index profile was recently proposed in [13]. The present solution offers one additional degree of freedom, not to mention the four-fold profile multiplication. Moreover, with little additional effort, one can build even more complex solvable profiles by iterating further the PROFIDT process.

### *5.4. One single PROFIDT with seed-potential $V_{[0]} = \beta \neq 0$ and transformation parameter $p_1 = 0$*

The case that may be of most benefit in the realm of optical thin layers because potentially leading to numerous practical applications is obtained after one single PROFIDT, starting from a constant, non-zero, seed-potential ($V_{[0]} = \beta \neq 0$) and choosing a zero transformation-parameter $p_1 = 0$. The admittance profiles and the wave-functions stemming from this combination are summarized in Table 1, cases c7-c8 for $\beta$>0 and case c10 for $\beta$<0. The new potential-function takes the form of a shifted Darboux-Pöschl-Teller potential whereas the transformed wave-function $\psi_{[1]}(\xi, k_0)$ reads (see Table 1 for the symbols):

$$\psi_{[1]} \propto \begin{bmatrix} K_{[1]} \\ P_{[1]} \end{bmatrix} \equiv \begin{bmatrix} \left(i\sqrt{k_0^2-\beta}-\sigma_{[0,p_1]}\right)C^{(\beta)} \\ \left(-i\sqrt{k_0^2-\beta}-\sigma_{[0,p_1]}\right)S^{(\beta)} \end{bmatrix}. \tag{67}$$

Three classes of profiles of the transformed admittance $s_{[1]}(\xi)=\eta_{[1]}^{\pm1/2}(\xi)$ are obtained, which are defined with *four* free-parameters. The properties of these profiles were already studied in [36] and a clear advantage was shown for the following class, which corresponds to $\beta>0$ and to the log-derivative $\sigma_{[0,p_1]}=\xi_c^{-1}\tanh(\hat{\xi})$ (case c7):

$$s_{[1]}(\xi)=\eta_{[1]}^{\pm1/2}(\xi)\propto\begin{bmatrix} \operatorname{sech}(\hat{\xi}) \\ \sinh(\hat{\xi})+\hat{\xi}\operatorname{sech}(\hat{\xi}) \end{bmatrix}; \qquad \begin{cases} \hat{\xi}\equiv\xi/\xi_c+\gamma \\ \xi_c\equiv|\beta|^{-1/2} \end{cases} \tag{68}$$

where $\operatorname{sech}=1/\cosh$ is the hyperbolic secant function (these profiles were dubbed "$\operatorname{sech}(\hat{\xi})$-type profiles"). As a matter of fact, these profiles present very interesting properties: 1) it was conjectured that they are perfectly *versatile*, that is, whatever the specified values for the property $s$ and for its slope at both ends of a given layer, there exists a (unique) set of four parameters ($\xi_c$, $\gamma$ and the two constant factors of the LC in eq. (68)) allowing the profile to satisfy the previous four conditions; 2) they satisfy the *parsimony principle* since they are defined with *no more parameters* than the number of specified conditions, namely *four*; 3) the profiles in eq. (68) and the related wave-functions $\psi_{[1]}$ in eq. (67) are exclusively based on elementary functions (basically exponential functions).

The transfer matrix corresponding to the $\operatorname{sech}(\hat{\xi})$-type profiles is easily computed by substituting in eq. (A2)-(A3) the expressions for $K_{[1]}$ and $P_{[1]}$ in eq. (67).

Let us mention that a class of $\operatorname{sech}(\alpha z)$-type profiles but with *only three parameters* was recently used for analysing mode-propagation and high-resolution imaging through a secant-index planar waveguide [57]. The LC described in eq. (68), though again with *only three parameters* (namely with $\gamma$ =0), was also elaborated in [58] as the generic solution of the Schrödinger equation for the one-soliton potential and corresponding to its single discrete spectrum level.

Skipping the details of the inverse-Liouville transformation, we directly provide the following results, valid for normal incidence or for oblique incidence with TE-waves only. The link between the $\xi$-scale spanning over $[0,\xi_1]$ and the $z$-scale spanning over $[0,z_1]$ is defined by:

$$\frac{z(\xi)}{\xi_c} = f(\xi) - f(0) \tag{69}$$

where the function $f(\xi)$ can take two forms: $f_E(\xi)$ is for $\langle E \rangle$-form profiles and $f_H(\xi)$ is for $\langle H \rangle$-form profiles:

$$f_E(\xi) = \begin{cases} \dfrac{-1}{2A_D} \dfrac{\operatorname{sech}(\hat{\xi})}{n_{[1]}^{*\,1/2}(\xi)} & \text{if} \quad A_D \neq 0 \\ \dfrac{1}{2A_B} \dfrac{\sinh(\hat{\xi}) + \hat{\xi}\operatorname{sech}(\hat{\xi})}{n_{[1]}^{*\,1/2}(\xi)} & \text{if} \quad A_B \neq 0, \end{cases} \tag{70}$$

$$f_H(\xi) = A_B{}^2 \tanh(\hat{\xi}) + A_D{}^2 \left[ \frac{1}{4}\sinh(2\hat{\xi}) - \frac{\hat{\xi}}{2} + \hat{\xi}^2 \tanh(\hat{\xi}) \right] + 2A_B A_D \hat{\xi} \tanh(\hat{\xi}), \tag{71}$$

where $A_B$ and $A_D$ are the constant coefficients multiplying the LIS in the definition of $n_{[1]}^{*\,\pm 1/2}(\xi)$ (not $\eta_{[1]}^{\pm 1/2}(\xi)$, see eq. (68) and (C.9)).

A complementary illustration of the "flexibility" of the $\operatorname{sech}(\hat{\xi})$-type profiles is given in Figure 13 where nine different combinations of right and left slopes were specified for a given value of the right-to-left index ratio. For each set of specifications, there are always two solutions: one of $\langle E \rangle$-form and another one of $\langle H \rangle$-form. This opens up a very promising new avenue for building synthetic solvable profiles that are not only continuous but also smooth, namely with a continuous first derivative and maybe even second derivative, as illustrated next.

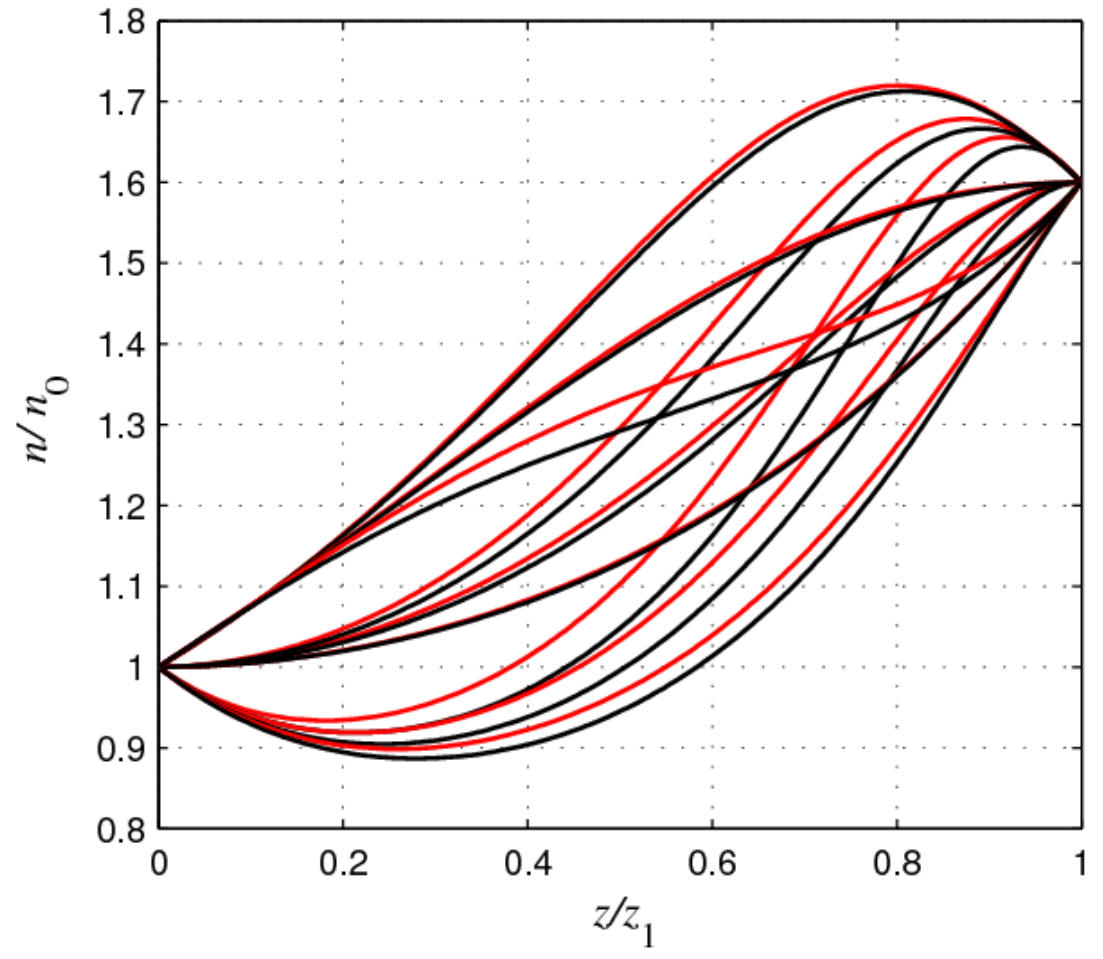

Figure 13. Solvable refractive-index of $\mathrm{sech}(\hat{\xi})$-type (eq. (68); case c7). The ratio of right-to-left refractive indexes is set to 1.6. The normalized slope was set to -1, 0 or 1 at each edge, which gives nine profiles of $\langle E \rangle$-form (in black) and nine profiles of $\langle H \rangle$-form (in red). The profiles are represented vs. the normalized physical depth $z/z_1$, i.e. after an inverse-Liouville transformation assuming normal incidence (see eq. (69)-(71)).

## 6. Applications of the $\mathrm{sech}(\hat{\xi})$-type profiles

### *6.1. Matching layers*

A first application is the design of *matching layers*. Graded matching layers are aimed at reducing the reflectance induced by the refractive-index jump otherwise present at the interface between two dissimilar media. One of the most popular designs, namely the *quintic* layer, follows a 5$^{th}$-degree polynomial profile [59]. The $\mathrm{sech}(\hat{\xi})$-type profiles with horizontal slopes at both ends actually present an interesting alternative. Figure 14 allows comparing a quintic profile and the new $\mathrm{sech}(\hat{\xi})$-type profiles of $\langle E \rangle$-form and of $\langle H \rangle$-form when matching two media of refractive indices 1.6 and 2.4. The latter two profiles are close one to the other (remember that, depending on the left and right slope combination, the $\langle E \rangle$-form and $\langle H \rangle$-form may be more or less separated, refer to Figure 13).

The related reflectance spectra are reported in Figure 15 vs. the reduced wavelength $\tilde{\lambda} = \lambda / n_0 z_1$ where $z_1$ is the physical-thickness of the matching-layer and $n_0$ is the index of the incident medium. The spectra of the $\langle E \rangle$-form and $\langle H \rangle$-form profiles are hardly distinguishable since the two profiles are very close too. Notice that if we normalize the reduced wavelength by $\xi_1$ instead, where $\xi_1$ is the optical-thickness of the matching-layer, the two $\text{sech}(\hat{\xi})$-type spectra perfectly overlap (*spectrally indistinguishable profiles*). For $\tilde{\lambda}$ larger than about 4-5, the reflectance is close to an asymptotic limit which actually corresponds to the Fresnel reflectance for non-matched media, namely 0.053 in the present case. The reduced wavelength needs to be less than a couple of units for the matching layers to be really efficient. Down to a reduced wavelength of about 0.9, the reflectance reduction is better with the $\text{sech}(\hat{\xi})$-type profiles than with the quintic profile. For a still decreasing wavelength, the reflectance drop is *globally* faster with the quintic profile than with the two others, even though, at specific wavelengths, the $\text{sech}(\hat{\xi})$-type profiles show local minima with a very low reflectance. Hence, the respective advantages are more or less balanced. Nevertheless, there is a clear computational advantage in using the $\text{sech}(\hat{\xi})$-type profiles over the quintic one since the optical response of the former ones is direct and exact. The transfer matrix described in Annex A indeed yields an exact modelling of the optical response of the *whole* matching layer. There is no need for discretizing this layer into thin homogenous sub-layers contrarily to the quintic layer (or any other profile that does not lead to an exact analytical EM solution). A more extensive comparison with other solutions from the literature for the purpose of index-matching will be the subject of a forthcoming paper.

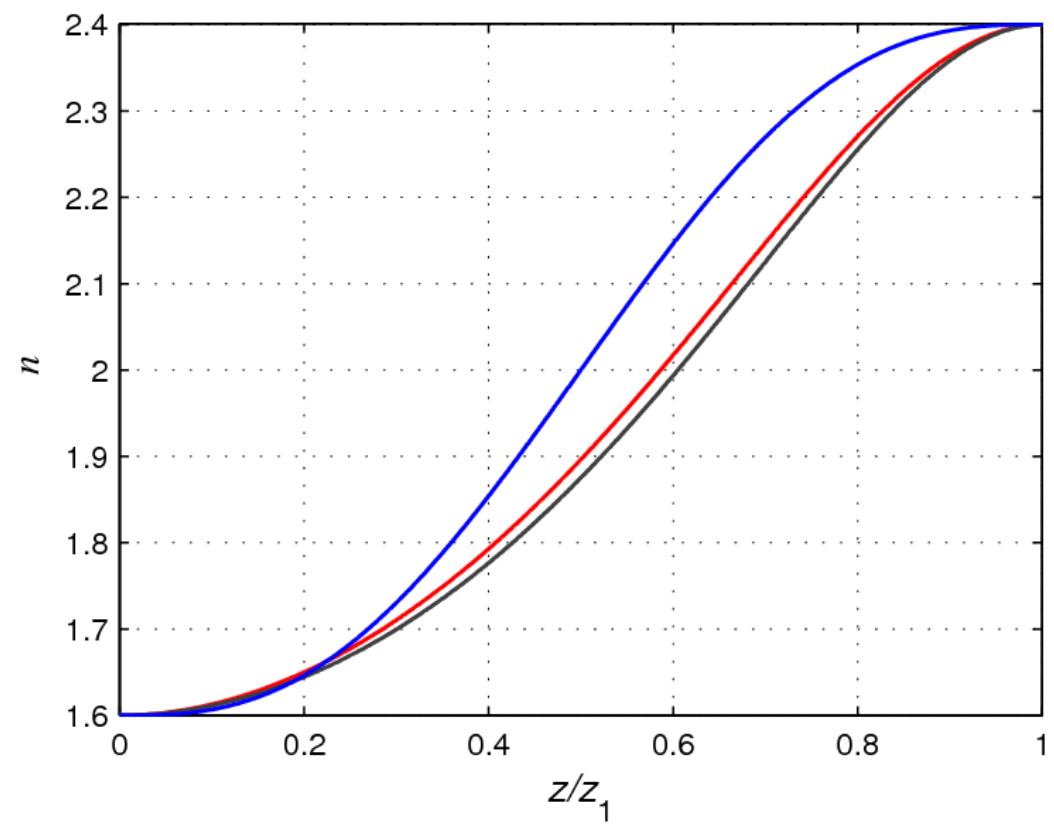

Figure 14. $\mathrm{Sech}\left(\hat{\xi}\right)$-type profiles of $\langle E \rangle$-form (in black) and of $\langle H \rangle$-form (in red) compared to a quintic profile [59] (in blue) used as matching-layers between two media of indexes $n_0$=1.6 and $n_1$=2.4. The depth $z$ is normalized by the layer thickness $z_1$.

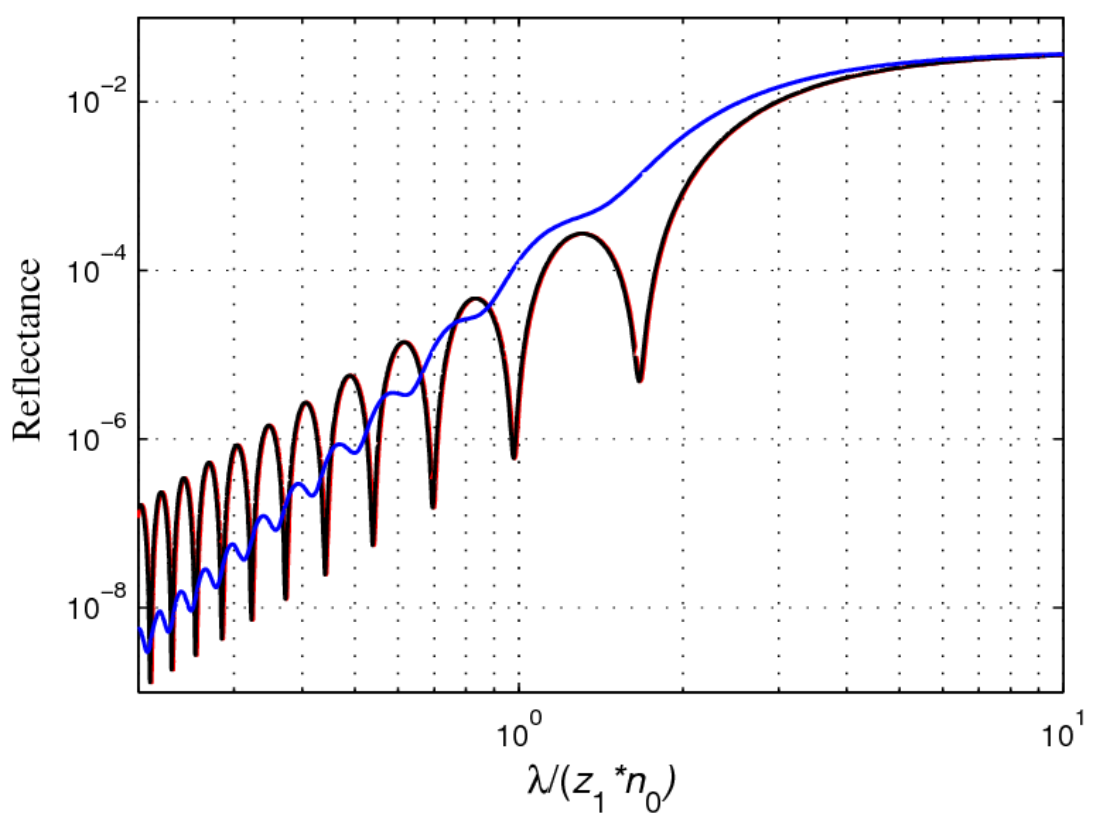

Figure 15. Reflectance spectra of the three matching-layers in Figure 14 with respect to the reduced wavelength of the incoming light (same colors). The free-space wavelength $\lambda$ is here divided by the coating thickness $z_1$ and by the incident-medium index. The spectra of $\langle E \rangle$-form and of $\langle H \rangle$-form are almost indistinguishable.

### *6.2. Profile synthesis by assembling* $\mathrm{sech}\left(\hat{\xi}\right)$*-type profiles*

A *rising* $\mathrm{sech}\left(\hat{\xi}\right)$-type profile of zero-slope at both ends like in Figure 14 can be joined to a symmetrically *falling* profile to produce an elementary pattern that can be later replicated for building a synthetic *periodic* (solvable) profile. Interestingly, the profiles produced thereby are continuous up to the *second* derivative at the nodes, unlike those resulting from assembling concave and convex profiles of “fundamental” type (see §4 and Figure 2; the curvature of the latter synthetic profile is indeed discontinuous at the nodes [10]).

One may argue that the $\mathrm{sech}(\hat{\xi})$-type profiles are not quite symmetric: the curvature in Figure 14 is indeed higher in the higher part than in the lower part. However they are getting more and more symmetric as the index contrast $n_1/n_0$ is approaching unity. They actually resemble more and more to a sinusoidal profile (not reported here). As a consequence, sinusoidal index-modulations of small magnitude, as encountered for example in Bragg filters, could be well modelled by a set of joined $\mathrm{sech}(\hat{\xi})$-type profiles.

Combining rising and falling $\mathrm{sech}(\hat{\xi})$-type profiles that have a horizontal slope at both ends but that are *not symmetric* would yield synthetic profiles that are continuous at the nodes up to the *first derivative only*. Yet, this offers an interesting opportunity for analytically synthetizing a huge variety of optical devices as shortly exemplified bellow. As a matter of fact, from one elementary $\mathrm{sech}(\hat{\xi})$-type profile to the next, one can alter the amplitude, the width, or both. This yields tapered (i.e. apodized) or chirped pseudo-periodic profiles or a combination of both.

Before going further, let us mention that the replication of an elementary pattern encompassing one or more pairs of rising and falling $\mathrm{sech}(\hat{\xi})$-type profiles would give rise to a periodic pattern that may be continuous up to the first derivative. This offers a new modelling tool for complex 1D photonic crystals. Indeed, this entails a new type of solutions for the Hill’s equation (in contrast to those presented in [14]-[18] which are based on trigonometric functions, the new one would be based on hyperbolic and linear functions of the optical-thickness) and a new avenue for studying the photonic bandgaps of periodic smooth multilayers [60].

The scattering properties (reflectance, transmittance, phase delay, etc...) of a composite profile made of a succession of elementary $\mathrm{sech}(\hat{\xi})$-type profiles, eventually followed by a semi-infinite profile of $\tanh^{\pm 1}(\hat{\xi})$-type, are easily obtained by the transfer-matrix method, namely by multiplying the transfer-matrices related to each elementary profile. As a consequence, the synthetized profile can also be regarded as *solvable*.

### *6.3. Rugate filters*

Rugate filters (or gradient-index filters) offer an efficient solution for producing notch filters deprived from ripples and side-lobes (see [61], [62], and a few more recent references: [63]-[68]). Tailored refractive indices can be reached by Ion Beam Sputtering deposition: with Si and Ta targets, say, a mixture film is formed yielding a continuous distribution of the refractive index between the values corresponding to the pure materials $SiO_2$ and $Ta_2O_5$, i.e. 1.45 and 2.24 at 500 nm [69]. Atomic layer deposition is another possibility: by regulating the thickness ratio of $TiO_2$ and $Al_2O_3$ in a nanoscale layer, the refractive index is tailored between the refractive indices of the two materials, viz. 2.42 and 1.61 at 550 nm [70]. Other techniques for getting graded-index filters with varying concentration of $SiO_2$ and $TiO_2$ are plasma enhanced- or ion beam induced chemical vapour deposition [71], [72].

Rugate-filter synthesis, namely the design of a refractive-index profile for matching a particular reflectance/transmittance curve, is out of scope in this paper; we will merely present an illustration of the direct approach, it means the calculation of the transmittance of a rugate filter that is modeled by a (smooth) sequence of $\operatorname{sech}\left(\hat{\xi}\right)$-type profiles. Figure 16 illustrates such a structure: the aim is to build a notch filter with a stop-band at 532 nm laid over a glass substrate with index $n_s$ =1.52 (we will not consider here the index step at the front surface; with an air interface, a supplementary antireflection coating or a matching layer would be necessary). Thirty pairs of alternatively rising and falling $\operatorname{sech}\left(\hat{\xi}\right)$-type profiles with horizontal slopes and 133 nm optical-thickness each were joined together. The mean index of the rugate is 2 and the nominal min/max values are 1.8 and 2.2. A supplementary $\operatorname{sech}\left(\hat{\xi}\right)$-type profile of optical-thickness 600 nm is added for achieving index-matching with the substrate at right. For reducing the side-lobes in the transmittance spectrum, the use of apodization is of great help [61], [62]. We did not optimize the apodization shape, we merely noticed that a centered square-sinus function applied on a stretched length scale produced quite satisfactory results; see the transmittance spectrum in black in Figure 17. We also considered two other rugates made of 50 and 70 pseudo-periods (resp. blue and red curves in Figure 17). Increasing the number of pseudo-periods improves the filter performance: the optical density increases from 1.9 to 3.6 and then 5.2. With just 30 periods, the rugate produces two small side-lobes where the transmittance drops to 0.979 and 0.969. These side-lobes are highly attenuated by adding 20 periods: the transmittance in the residual side-lobes is then higher than 0.993. In all three cases, the residual ripples are quite small over a very large bandwidth.

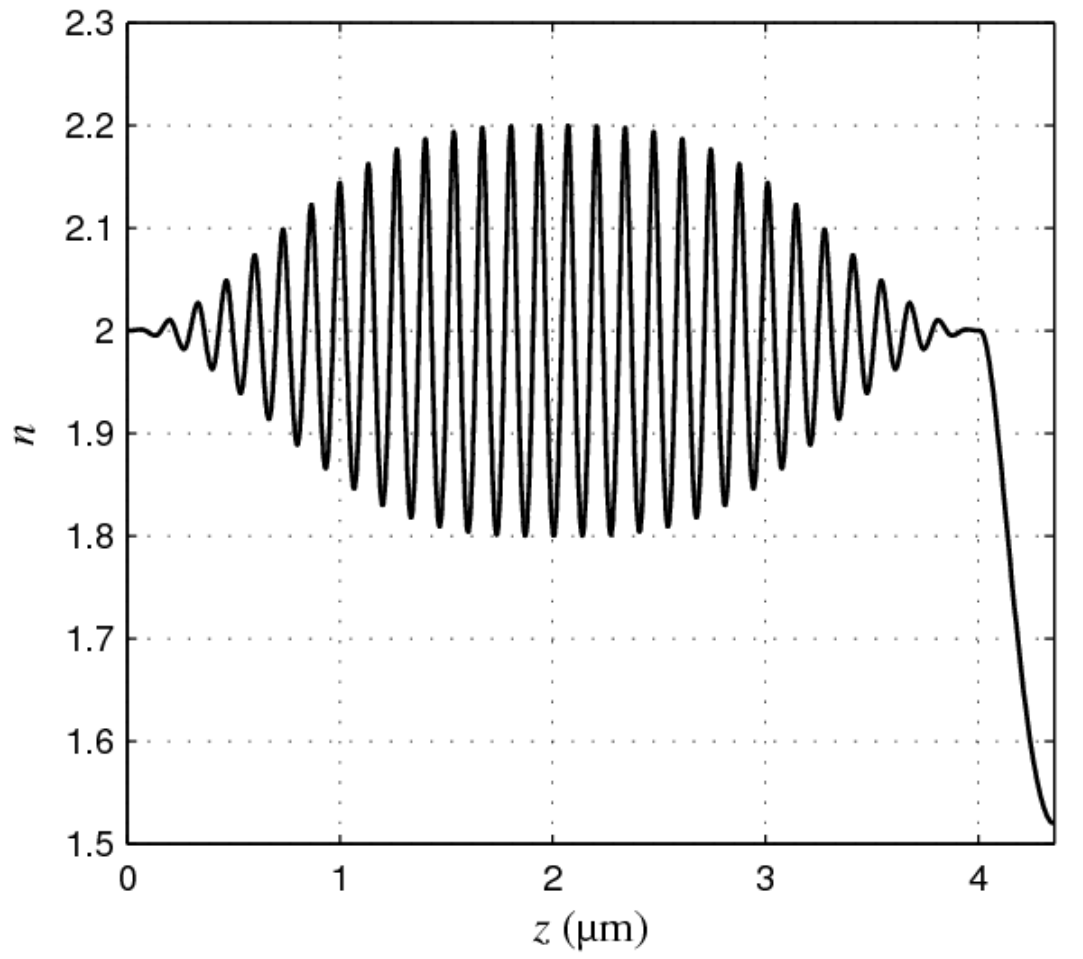

Figure 16. Rugate filter (notch filter) obtained by joining 60 $\mathrm{sech}\left(\hat{\xi}\right)$-type profiles of $\langle E \rangle$-form resulting in 30 periods of 266 nm optical-thickness each (mean index: 2; peak-to-peak amplitude: 0.4; square-sinus apodization). A matching-layer with 600 nm optical-thickness is added at right for a smooth transition between the mean index (2) and the substrate index (1.52).

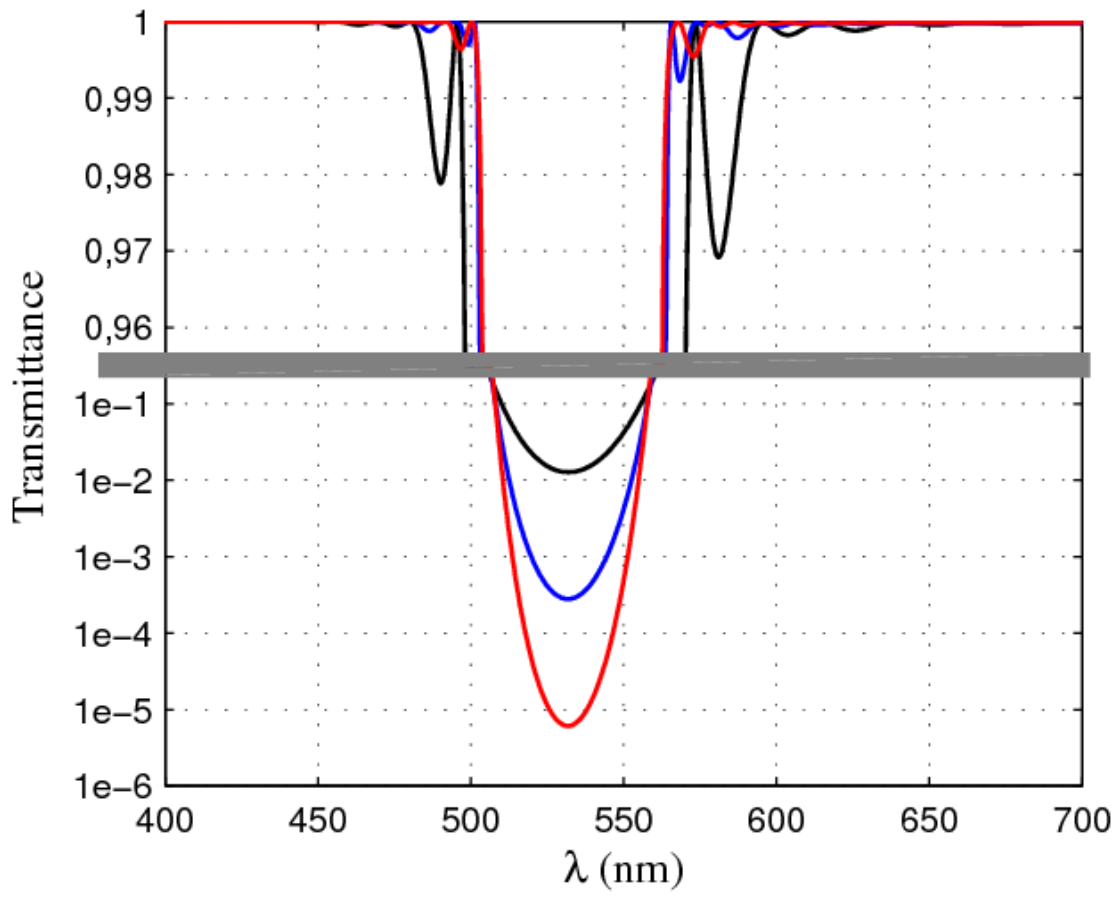

Figure 17. Transmittance spectra of three notch-filters according to the scheme described in Figure 16 (532 nm central wavelength). They differ in the number of periods of the rugate: 30 periods, 4.35 µm total thickness (in black – this case is represented in Figure 16); 50 periods, 7.03 µm total thickness (in blue); 70 periods, 9.70 µm total thickness (in red). The transmittance scale is split into two parts for zooming independently in the high-transmission region and in the low- transmission region (stop-band); a linear scale is used for the upper part and a logarithmic scale is used for the lower part.

### *6.4. Chirped reflectors*

As a last example let us consider the case of optical devices based on a modulation of the refractive-index with a slow variation of the modulation period along the filter length (chirp). Chirped reflectors are widely used in the field of multilayer coatings aimed at ultrafast applications (femtosecond lasers). Pulse compression and more generally phase and dispersion compensation requires controlling accurately and over a wide bandwidth the group-delay (GD) and the group-delay-dispersion (GDD), which is the first (resp. the second) derivative of the phase upon reflection with respect to the angular frequency. This can be achieved by implementing so-called chirped mirrors [73]-[76], chirped fiber Bragg gratings [77], [78] or chirped volume Bragg gratings [79]. Chirped Bragg reflectors can also be implemented in a laser cavity for widening the resonant modes. Also, chirped photonic crystals can reflect light over a larger bandwidth than do crystals with regular lattice; a series of such examples can be found in nature in biological reflectors, e.g. certain beetle elytra and butterfly eyes [80].

When designing a chirped mirror for laser pulse compression, an initial *continuous* refractive-index profile may be estimated by applying the classical Fourier-transform approach, based on the mirror reflectance expectation [73]. However, most often the final configuration takes the form of a complex multilayer stacking made of a pair of high and low index materials. This final design (i.e. the sequence of thickness values of the multilayer) is the result of a computer optimization which aims at reaching simultaneously a high reflectivity, a high negative GDD deprived of oscillations and a low sensitivity to design and fabrication errors [73]-[76]. The purpose of the negative GDD is to compensate the (always) positive GDD introduced by the optical materials in the laser system (solid-state gain media, air ...). Furthermore, ultrashort pulse compression requires dispersion controlling devices spanning over a wide spectral range: for example, with a Ti:sapphire laser (800 nm), the dispersive mirror should cover *one optical octave* in view of compressing the pulse down to about 4 fs. It is well recognized that impedance-matching problems, especially at the interface between the mirror and the ambient medium are at the origin of deviations from the desired dispersion characteristics, most often in the form of spurious oscillations.

The following simulation exercise focuses on the wideband performances of a *chirped mirror* of *rugate-type* targeted to that task. Among the different possible configurations aimed at reducing the unwanted interferences between rays reflected at the front interface and rays reflected deeper in the mirror coating, we selected the tilted-front-interface concept (TFI) [75], [81]. In this embodiment a thin glass-wedge is attached to the coated glass-substrate. The remaining reflections at the air-wedge interface don't disturb the dispersion properties anymore [75], [81]. Hence we will focus on the index-matching problems between the rugate and the surrounding glass material.

Again, a series of $\operatorname{sech}\left(\hat{\xi}\right)$-type profiles were assembled for building a synthetic profile as represented in Figure 18. This time, the height, the width and the mean level of the elementary profiles were all three slowly changed from one end of the mirror to the other. The graded index evolves between the values corresponding to glass (1.52), $SiO_2$ as the low-index material (1.45) and $TiO_2$ as the high-index material (2.42). The rugate reported in Figure 18 actually realizes a *3-fold smooth transition:* the spatial frequency, the local minima and the local maxima. The optical-thickness related to each pseudo-period smoothly increases with a central value of 400 nm which corresponds to a Bragg wavelength of 800 nm. In essence, this rugate model can be seen as a transposition and a generalization in the "rugate world" of the double-chirped mirror structure [82] and the dual-adiabatic-matching structure [83]. This design can accommodate through adiabatic matching any combination of front and substrate media provided their index is in the range of values that can be reached through the manufacturing techniques mentioned before. Obviously, the same approach could be used for modelling chirped fiber Bragg gratings, i.e. devices with substantially lower peak-to-peak index modulations.

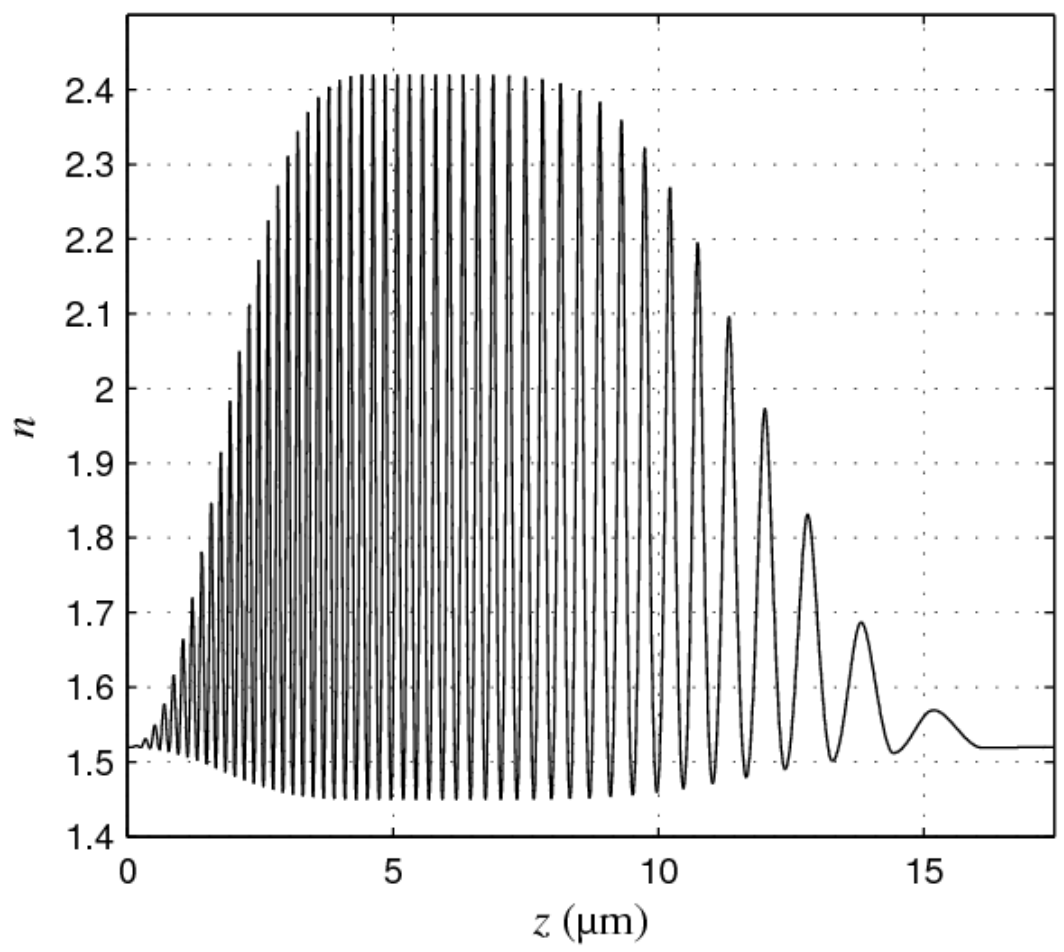

Figure 18. Modelling of a chirped mirror for ultrafast laser pulse compression (800 nm range). The rugate is deposited on a glass substrate (at right) and joined to a glass wedge (at left) which acts as the incident medium. Rugate made of 50 pseudo-periods based on pairs of $\mathrm{sech}(\hat{\xi})$-type profiles ($\langle E \rangle$-form) with progressively rising optical-thickness.

The reflectance spectrum of the chirped mirror in Figure 18 is reported in Figure 19. The rugate is made of 50 pseudo-periods which amounts to a physical thickness of about 16 μm. The group delay and group-delay-dispersion are plotted in Figure 20. Notice that the results are nearly undistinguishable, whether all $\mathrm{sech}(\hat{\xi})$-type elementary profiles are of $\langle E \rangle$-form (in black) or $\langle H \rangle$-form (in red).

After very simple adjustments of the apodization and the implementation of a slightly non-linear chirp we could obtain quite interesting results, without the need for involved optimization procedures, as usually required in the case of multilayer structures [73]-[76]. The GD curve is actually very smooth while being close to linear. Moreover, only slight oscillations can be noticed on the GDD curve. While setting the reflectance threshold at 0.95 and allowing an excursion of 30 $fs^2$ for GDD (with respect to the mean value), we infer that the designed mirror has a *one-octave range* (i.e. from 600 to 1270 nm). In this range the GDD shows only minimal variations of ±8 $fs^2$ around a reasonably high value (in magnitude) of -50 $fs^2$. Still better results could be obtained by further optimizing the rugate chirp and the apodization. Easy changes can also be introduced for taking into account the dispersion of the low- and high-index materials used for manufacturing the rugate. The results of these modifications together with a sensitivity analysis to design errors (i.e. to small perturbations of chirp and amplitude) will be presented in a future paper.

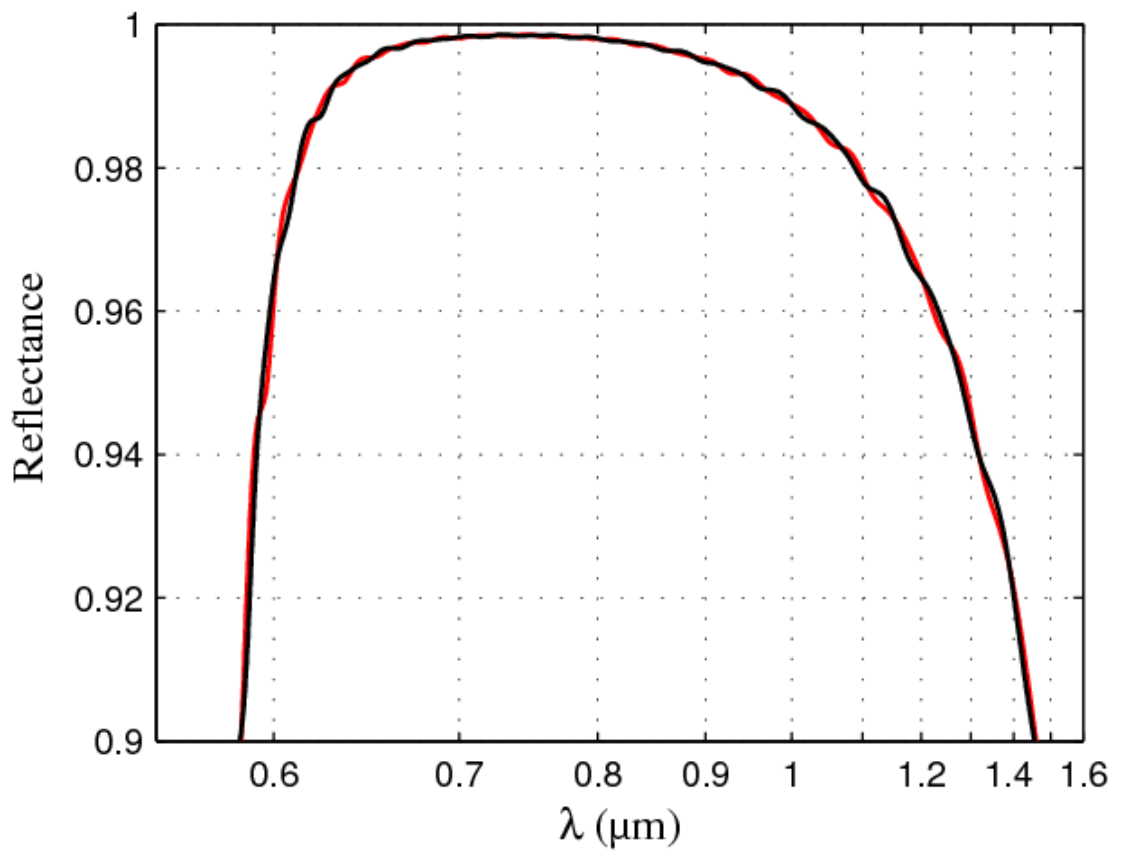

Figure 19. Reflectance spectrum of the chirped-mirror design in Figure 18 (the result for a rugate built with $\langle E \rangle$-form profiles, resp. $\langle H \rangle$-form profiles, is in black, resp. in red).

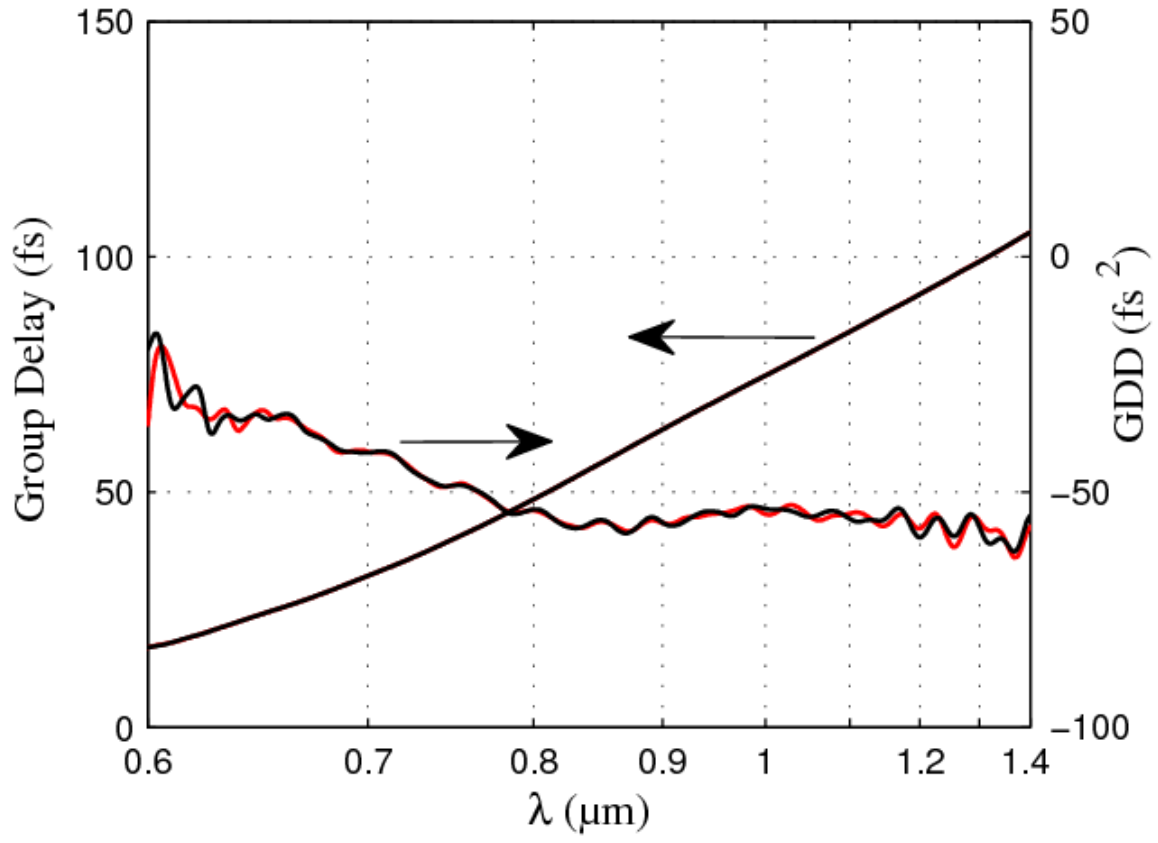

Figure 20. Same as in Figure 19 for the group delay and the group-delay-dispersion (GDD)

## 7. Discussion and concluding remarks

### *7.1. Darboux transformations in $\xi-$space vs in $z-$space*

As far as we know, the Darboux transformation (or the related supersymmetry (SUSY) optics theory), when applied to the equations of EM-wave propagation in heterogeneous media, has always been implemented in the *physical-coordinate space* (see e.g. [24], [27]-[32]). The main reason is that in the $z-$space, the electric-field in non-magnetic materials and TE-mode can be easily expressed in the form of a time-independent Schrödinger equation (after an intermediate step featuring a Helmholtz equation - refer to eq. (4) with $m=+1$ - or a Dirac equation resulting from the application of the coupled-mode theory [24]). However, the equation of the magnetic-field for the TM-mode involves a weighting inside the $z-$derivative by a non-constant function, namely $1/\varepsilon(z)$, or equivalently $1/n^2(z)$ (refer to eq. (5) with $m=-1$). For solving this difficulty, a (partial) Liouville transformation was implemented in [29], [30] involving a dependent-variable transformation but no independent-variable transformation. The Darboux transformation (or the equivalent supersymmetry process) was then performed on the resulting stationary Schrödinger equation for building sequences of solvable profiles related to an *effective* permittivity. Thereafter, accessing to the *actual* permittivity requires still solving an auxiliary 2nd order *non-linear* equation, which cannot be performed analytically. On the other side, in [32], both permittivity and permeability were assumed to be (1D) space-dependent. The consequence is that both equations for TE and TM modes suffer from the difficulty described just before. The authors pursued the analysis by applying a (partial) Liouville transformation and considering only the very special case of a constant refractive-index, i.e. $\varepsilon(z)\mu(z)$ =cst.

The Liouville transformation in [29], [30], [32] involved no independent-variable transformation contrary to what was done in the present paper (see eq. (6)). The consequence is that the "eigenparameters" (or the state energies) in the resulting Schrödinger equations are of different nature. When working in the $z-$space and after a scaling by $k_0$, the "eigenparameter" comes out to be $-I_a{}^2$. The equation is then suitable for addressing the *waveguide properties* of the considered optical device at a *given wavenumber* ($I_a{}^2$ corresponds to $\beta/k_0^2$ where $\beta$ is the well-known *propagation constant*). Thereafter, Darboux or SUSY transformations allow generating new potentials (namely new permittivity - or effective permittivity - profiles) which are valid for the *wavenumber* under consideration. By contrast, when working in the $\xi-$space (i.e. after a *full* Liouville transformation), the "eigenparameter" comes out to be $k_0^2$ (see eq. (12) (13)). The equations are then suitable for addressing the *spectral properties* of a graded material when submitted to an EM plane-wave at *given incidence*. The interesting point is that both TE and TM fields then share the *same* Schrödinger equation; the potential-function is simply the relative curvature (in the optical-thickness $\xi$ – space) of the square-root of the tilted admittance (for the transformed $E$ -field) or its reciprocal (for the transformed $H$ -field). The classical Darboux transformation can then be applied equally well to both polarization modes. Moreover, the core of the present work lies in the fact that the tilted admittance (or its reciprocal) satisfies the *same* Schrödinger equation as do the $E$ -field (resp. the $H$ -field), except that the "eigenparameter" is set to 0. As such, the PROFIDT methodology that was devised in [36] for solving thermal-diffusion problems allows building presently sequences of solvable admittance profiles along with the related EM-field solutions. Similarly, at each step, two families of profile solutions are obtained: the first one stems from the $E$ -field equation and the other one from the $H$ -field equation. The only remaining difficulty stands in the inverse-Liouville transformation, it means in establishing an analytic closed-form relation between the optical-thickness $\xi$ and the physical-coordinate $z$ . This operation involves a quadrature, which, in the worst case, has to be performed numerically. Yet, for many important cases (TE-mode in non-magnetic, lossless media), we provided analytical expressions for performing this inverse transformation (though, sometimes in implicit form only).

Recall that the *refractive-index* profiles and the related scattering properties (i.e. reflectance/transmittance spectra) that may be inferred from the *admittance* profiles are jointly bound to a *chosen incidence.*

### *7.2. PROFIDT-generated solvable profiles*

All the examples described in this paper originate from a *constant* seed-potential. The admittance profiles and EM-field solutions related to this trivial potential (the so-called fundamental solutions) are already quite interesting, some of them having not been published yet. Starting the PROFIDT process from a constant potential has the advantage of producing solutions that are exclusively based on *elementary functions (basically exponential),* whatever the PROFIDT order. We also established that in some circumstances the admittance profiles of $\langle E \rangle$ -form and $\langle H \rangle$ -form are *spectrally indistinguishable* (i.e. they produce exactly the same reflectance amplitude spectrum).

Solvable admittance (or refractive-index) profiles with progressively more sophisticated shapes can be obtained by iterating further the PROFIDT process or by restarting from other, more complex, seed-potentials. With a *constant* seed-potential, the profiles generated after a cascade of $J$ Darboux transformations are expressed with as much as $3+2J$ free-parameters. The analytical solutions produced in this way are getting inexorably richer and richer, though computationally easily tractable, which is a good point. There is however one drawback to mention: all but two parameters are non-linear. Furthermore, the few cases we could explore also showed severe parameter-correlations. It means that if one wanted to fit a prescribed admittance profile with a solvable profile having a high number of parameters, one would probably face thorny problems. We however found a remedy for this difficulty through the use of a particular class of solvable profiles, the $\operatorname{sech}(\hat{\xi})$-type profiles.

The 4-parameter $\operatorname{sech}(\hat{\xi})$-type admittance (or refractive-index) profiles that were here described for the first time (although they share the same shape as the $\operatorname{sech}(\hat{\xi})$-type thermal-effusivity profiles), present an outstanding interest due to their extreme versatility and flexibility, despite the reduced number of constitutive parameters. A few examples were presented, all based on a sub-class of $\operatorname{sech}(\hat{\xi})$-type with horizontal end-slopes. It was clearly shown that these specific profiles, together with the related EM-field solutions, could be advantageously used as *high-level analytical tools* for modelling matching layers, rugate filters, Bragg filters and chirped mirrors, among others. Quite probably they could also be successfully used for modelling 1D-graded photonic crystals [4], [5], [7], [14]-[18], aperiodic quasi-crystals [84] and alike optical devices.

The optical scattering properties of a graded-index profile are most often evaluated with the classical transfer-matrix method based on a *piecewise constant approximation* of the profile (it is exact only in the asymptotic sense). It requires selecting carefully the thicknesses $\Delta z_i$ of the sublayers with respect to the wavelength under consideration: as a rule of thumb, for good numerical results, any optical-thickness $n_i \Delta z_i$ should be less than about $\lambda/60$. As such, for the three rugates whose transmittance spectra are presented in Figure 17, the minimum number of homogeneous sublayers should be about 1300, resp. 2100 and 2900 (based on $\lambda_{\min}$ = 400 µm and $\xi_{total}$ = 8.6, resp. 13.9 and 19.2 µm), which means as many transfer-matrices to compute and multiply together. By contrast, only 61, resp. 101 and 141 (specific) transfer-matrices (two per period plus one for the matching layer at right) were necessary when using $\operatorname{sech}(\hat{\xi})$-type profiles, which is about 20 times less than with the classical method. In addition to solving (or at least mitigating) the round-off error accumulation problem, the new method eliminates the worry about the minimal-discretization criterion.

In contrast to similar techniques based on "high-order" transfer-matrix manipulations (e.g. [8], [9]), the present computation involves only calls to *exponential functions*.

When addressing *modulated* index profiles, other classical methods are the perturbation-based methods among them the coupled-mode method (see for example [47], [60], [78], [80]). A systematic starting hypothesis is that the index variations are of small amplitude. Throughout the theory we presented in this paper, there is no such limitation: the solvable profiles are *arbitrarily scalable* and the corresponding *EM-field analytical solutions are exact*.

Furthermore, just like in the context of thermal-effusivity profile-synthesis [36], we propose to use $\operatorname{sech}(\hat{\xi})$-type profiles as elementary bricks for approximating an arbitrary admittance (or refractive-index) profile. More precisely, given a set of admittance nodes, *spline interpolation* could be performed by using $\operatorname{sech}(\hat{\xi})$-type profiles in lieu of the usual cubic polynomials. Mimicking the cubic-spline interpolation, the four parameters of each $\operatorname{sech}(\hat{\xi})$-type element will be adjusted so that the (smooth) synthetic admittance profile passes through all nodes while being continuous up to the second derivative at each internal node. The immense benefit of this approach is that the optical response of the whole ensemble is analytical; it simply entails the multiplication of the transfer matrices related to each spline segment. Furthermore it is still exact, with no limitation on the admittance-variation magnitude. The implementation of the dubbed *"*$\operatorname{sech}(\hat{\xi})$*-splines"* will be the object of a next paper.

Eventually, the PROFIDT method described in this paper is suitable for modelling other 1D-wave transport phenomena as briefly quoted in Annex D.

The present paper dealt with *real-valued positive* permittivity and permeability functions. A formal extension to *negative or complex-valued* functions should be feasible which would be interesting for modelling graded materials with loss and/or gain as well as graded metamaterials. Further work is needed for addressing this important topic.

## ANNEX A

The corner stone of the PROFIDT method is the *Darboux transformation* (DT), a differential transformation which, from the knowledge of the generic solution of a Schrödinger-like equation with given potential, provides the generic solution related to a *transformed potential* [25], [26], [51]. Since it can be applied recursively, we will introduce a subscript $j$ for specifying the step number in the process. The parameter will be given the symbol *p* like in [36], for ease. Given the equation:

$$\frac{d^2\psi}{d\xi^2} = \left(V_{[j]}(\xi) + p\right)\psi , \qquad \text{(A.1)}$$

for which we know a solution $\psi_{[j]}$ for any value of the constant parameter $p$. Let $\tilde{\psi}_{[j,p_{j+1}]}$ be a solution of eq. (A.1) for the particular value $p = p_{j+1}$ which is called the *transformation parameter*; $\tilde{\psi}_{[j,p_{j+1}]}$ is called the *transformation function*. Let $\sigma_{[j,p_{j+1}]}$ be its logarithmic derivative (the opposite is usually called the *superpotential*):

$$\sigma_{[j,p_{j+1}]} \equiv \left(\ln\left(\tilde{\psi}_{[j,p_{j+1}]}\right)\right)' = \frac{\tilde{\psi}_{[j,p_{j+1}]}'}{\tilde{\psi}_{[j,p_{j+1}]}} . \qquad \text{(A.2)}$$

The DT gives rise to a new equation:

$$\frac{d^2\psi}{d\xi^2} = \left(V_{[j+1]}(\xi) + p\right)\psi , \qquad \text{(A.3)}$$

with a transformed potential $V_{[j+1]}$ defined by:

$$V_{[j+1]} \equiv -V_{[j]} + 2\sigma^2[j, p_{j+1}] - 2p_{j+1}. \qquad \text{(A.4)}$$

Thereafter, if, for a given parameter $p = -k_0^2$, we know two linearly independent solutions (LIS) of eq. (A.1), $K_{[j]}$ and $P_{[j]}$ (and thus the corresponding generic solution $\psi_{[j]} \propto \left[K_{[j]} \quad P_{[j]}\right]^t$), the transformed functions $K_{[j+1]}$ and $P_{[j+1]}$ that are defined by:

$$\begin{bmatrix} K_{[j+1]}(\xi, k_0) \\ P_{[j+1]}(\xi, k_0) \end{bmatrix} \equiv \begin{bmatrix} K_{[j]}' - \sigma_{[j, p_{j+1}]} K_{[j]} \\ P_{[j]}' - \sigma_{[j, p_{j+1}]} P_{[j]} \end{bmatrix} \text{ if } k_0^2 \neq -p_{j+1}, \qquad \text{(A.5)}$$

$$\begin{bmatrix} K_{[j+1]}(\xi, k_0) \\ P_{[j+1]}(\xi, k_0) \end{bmatrix} \equiv \begin{bmatrix} \tilde{\psi}_{[j, p_{j+1}]}^{-1} \\ \tilde{\psi}_{[j, p_{j+1}]}^{-1} \int^{\xi} \tilde{\psi}_{[j, p_{j+1}]}^{2} d\xi' \end{bmatrix} \text{ if } k_0^2 = -p_{j+1}, \qquad \text{(A.6)}$$

are two LIS of eq. (A.3) for this same parameter $p = -k_0^2$ (as such, they define its general solution through: $\psi_{[j+1]} \propto \left[K_{[j+1]} \quad P_{[j+1]}\right]^t$).

The most interesting point is that, by virtue of the fact that the Liouville-transformed field $\psi(\xi, k_0)$ and the transformed admittance $s(\xi)$ satisfy two *homologous* Schrödinger equations (see eq. (15) and (16)), the Darboux transformation can thus provide simultaneously a new generic solution for profile $s(\xi)$ and the corresponding generic solution for the transformed EM-field $\psi(\xi, k_0)$. As a matter of fact, $s_{[j+1]}(\xi)$ can be simply obtained from $s_{[j]}(\xi)$ by repeating the process defined in eq. (A.5)-(A.6) but setting the parameter $-k_0^2$ to 0. As such, the generic solution of $s_{[j+1]}(\xi) \equiv \eta_{[j+1]}^{\pm 1/2}(\xi) \propto \left[B_{[j+1]}(\xi) \quad D_{[j+1]}(\xi)\right]^t$ at step $j+1$ will be given by:

$$\begin{bmatrix} B_{[j+1]}(\xi) \\ D_{[j+1]}(\xi) \end{bmatrix} \equiv \begin{bmatrix} B_{[j]}' - \sigma_{[j,p_{j+1}]} B_{[j]} \\ D_{[j]}' - \sigma_{[j,p_{j+1}]} D_{[j]} \end{bmatrix} \quad \text{if } p_{j+1} \neq 0, \tag{A.7}$$

$$\begin{bmatrix} B_{[j+1]}(\xi) \\ D_{[j+1]}(\xi) \end{bmatrix} \equiv \begin{bmatrix} \tilde{\psi}_{[j,0]}^{-1} \\ \tilde{\psi}_{[j,0]}^{-1} \int^{\xi} \tilde{\psi}_{[j,0]}^{2} d\xi' \end{bmatrix} \quad \text{if } p_{j+1} = 0. \tag{A.8}$$

The overall procedure consisting of a *transformation* of both the *field* $\psi_{[j]} \to \psi_{[j+1]}$ and the *property profile* $s_{[j]} \to s_{[j+1]}$ was called a *joint PROperty & FIeld Darboux Transformation* (PROFIDT) [36]. After each iteration, both new generic profile $s_{[j+1]}$ and new generic field $\psi_{[j+1]}$ are defined with *two* supplementary free-parameters compared to $s_{[j]}$, resp. $\psi_{[j]}$ (there is only *one* supplementary free-parameter when the transformation parameter $p_{j+1}$ is set to 0).

Since in the present analysis we restrict to real-valued, positive permittivity and permeability, (hence to real-valued, positive admittance), and since $s_{[j]}(\xi)$ is assumed to correspond to the square root of admittance (or the reciprocal), we have to impose that the solution $s_{[j]}(\xi)$ be real-valued and strictly positive. Let us mention that in the case of iterated PROFIDT process and if one is only interested in the final solution $s_{[J]}(\xi)$, the constraint actually applies only to the latter solution. All intermediate solutions $s_{[j]}(\xi)$, $j < J$, should be considered as simple mathematical objects and therefore not submitted to the previous constraint.

The interested reader may find in [36]:

- alternate expressions to eq. (A.6) and eq. (A.8) that are deprived of quadratures;
- synthetic expressions for the 2-fold PROFIDT;
- synthetic expressions for the $j$-order PROFIDT in the non-confluent case, i.e. when the successive transformation parameter values $p_1, \ldots, p_j$ are *pairwise different*.

**ANNEX B**

We provide the transfer matrices describing the EM-wave propagation between the optical-depths $\xi = \xi_0$ and $\xi = \xi_1$ inside a *graded* medium. The computation principle is well known: the first step is to express the vector $[E, H]^t$ by using eq. (23) and (24), where $s(\xi)$ and $\psi(\xi, k_0)$ should be expressed at $\xi = \xi_1$ (for $\psi$ we make use of the LC in eq. (21)). This linear system of equations is then solved for getting the two constants of the LC. They are then substituted in a system obtained by expressing the vector $[E, H]^t$ at $\xi = \xi_0$. This yields the linear relation in eq. (29) involving the transfer-matrix $\mathbf{M}_{\langle E \rangle}$. The four terms *A, B, C, D* of the matrix are obtained from:

$$\begin{bmatrix} A & B \\ C & D \end{bmatrix} = \frac{1}{\Delta}\begin{bmatrix} s_0^{-1} s_1 (G - \mu_1 I) & imk_0 s_0^{-1} s_1^{-1} I \\ imk_0^{-1} s_0 s_1 (-\mu_0 G - \mu_1 H + \mu_0 \mu_1 I + J) & s_0 s_1^{-1} (-H + \mu_0 I) \end{bmatrix} \quad , \qquad \mu_{0,1} = \frac{s_{0,1}'}{s_{0,1}} \tag{B.1}$$

where $G, H, I, J$ involve the values taken by the independent functions $K(\xi, k_0)$, $P(\xi, k_0)$ and by their derivatives at the optical depths $\xi = \xi_0$ and $\xi = \xi_1$:

$$\begin{aligned} G &= K_0 P_1' - K_1' P_0 \qquad & H &= K_0' P_1 - K_1 P_0' \\ I &= K_0 P_1 - K_1 P_0 \qquad & J &= K_0' P_1' - K_1' P_0' \end{aligned} \tag{B.2}$$

In (B.1) and (B.2) the subscripts 0 and 1 indicate that the corresponding functions are evaluated at $\xi = \xi_0$, resp. at $\xi = \xi_1$. $\Delta$ is the Wronskian of $K$ and $P$: $\Delta = W(K, P)$. Since it is constant ($K$ and $P$ are two independent solutions of a Schrödinger-type equation) it may be evaluated at $\xi = \xi_0$ or at $\xi = \xi_1$. Hence: . $\Delta = K_1 P_1' - K_1' P_1 = K_0 P_0' - K_0' P_0$.

By performing the same process, yet starting from eq.

(26) and (27) instead, we get another matrix $\mathbf{M}_{\langle H \rangle}$ that is now relevant to the solution obtained with the $\langle H \rangle$-form equation. It appears that the matrix $\mathbf{M}_{\langle H \rangle}$ is simply the result of a two-step circular permutation of the matrix $\mathbf{M}_{\langle E \rangle}$ (see eq. (30)). Furthermore, elementary algebra shows that both matrices are unimodular (i.e. their determinant is 1). In the case of *homogeneous* layer of admittance $\eta$, both transfer matrices $\mathbf{M}_{\langle E \rangle}$ and $\mathbf{M}_{\langle H \rangle}$ reduce to the well-known expression:

$$\mathbf{M} = \begin{bmatrix} \cos\delta & im\eta^{-1}\sin\delta \\ im\eta\sin\delta & \cos\delta \end{bmatrix}, \tag{B.3}$$

with $\delta = k_0 n \Delta z \cos\phi = k_0 \Delta\xi$ where $\phi$ is the constant incidence angle inside the layer of index $n$, thickness $\Delta z$ and optical thickness $\Delta\xi$. The reflectance $R$ of a stacking of $N$ layers sandwiched between a front material of admittance $\eta_a$ and a semi-infinite substrate of admittance $\eta_s$, is given by:

$$R = rr^*, \quad r = \frac{m\eta_a - Y}{m\eta_a + Y}, \quad Y = \frac{C + Dm\eta_s}{A + Bm\eta_s}, \tag{B.4}$$

where $r$ is the complex reflectance, $Y$ is the surface admittance of stacking+substrate and $A, B, C, D$ are the four terms of the global transfer matrix $\mathbf{M}$, i.e. as obtained by multiplying the $N$ matrices, from front edge to substrate: $\mathbf{M} = \mathbf{M}_1 \cdot \mathbf{M}_2 \cdot \ldots \mathbf{M}_N$. (notice that the value of the mode-coefficient, $m = \pm 1$, has finally no consequence on the complex reflectance value $r$; hence it could be discarded from eq. (B.1), (B.3) and (B.4)). Alternatively, the surface admittance $Y$ could be obtained as the final step $Y_0$ of a process starting from $Y_N = m\eta_s$ and defined recursively by:

$$Y_{k-1} = \frac{C_k + D_k Y_k}{A_k + B_k Y_k}, \tag{B.5}$$

where $A_k, B_k, C_k, D_k$ are the four terms of the transfer matrix $\mathbf{M}_k$ relatively to the k$^{\text{th}}$ layer.

## ANNEX C

This annex provides the detailed procedure for passing from the optical-depth space back to the physical-depth space, for non-magnetic materials, given a solvable pseudo-index profile $n^*(\xi)$, an invariant $I_a$ and a polarization mode (TE or TM).

(1) Normal incidence or TE-wave at oblique incidence.

Upon analytical evaluation of the integral $z(\xi)=\int_0^\xi n^{*-1}(u)du$ (see below for an alternative output), we get an expression linking the two space scales that we write as: $z = g(\xi)$. If the function $g$ is analytically invertible, i.e. $\xi = g^{-1}(z)$, the refractive-index profile in the physical-depth space is then obtained *explicitly* from:

$$n(z)=\sqrt{n^{*2}\left(g^{-1}(z)\right)+I_a{}^2}\,. \qquad \text{(C.1)}$$

If the function $g$ is not invertible, the index profile is defined *implicitly* by the system of equations:

$$\begin{cases} z = g(\xi) \\ n(\xi)=\sqrt{n^{*2}(\xi)+I_a^2}\,. \end{cases} \qquad \text{(C.2)}$$

It may happen that the function $n^* = f(\xi)$ itself is analytically invertible; in that case, the former implicit set reduces to a *single implicit* relation:

$$z = g\left(f^{-1}\left(\sqrt{n^2 - I_a^2}\right)\right). \qquad \text{(C.3)}$$

Let us mention that, for TE-mode, the integral at stake amounts to calculate:

$$z(\xi)=\int_0^{\xi} n^{*-1}(u)du=\begin{cases} y_0\int_0^{\xi} s^{-2}(u)du & \text{for } \langle E\rangle-\text{form} \quad \text{(C.4)}\\ y_0\int_0^{\xi} s^{2}(u)du & \text{for } \langle H\rangle-\text{form} \quad \text{(C.5)}\end{cases}$$

Hence it amounts to calculate the primitive of the square of a solution to a Schrödinger equation (or its inverse). As put forward in [36], the works in [48]-[50] provide two means for evaluating these primitives *without resorting to quadratures*, which may be helpful in the case of complicated expressions for $s^{\pm 2}(\xi)$. According to [36], the appropriate alternative expressions are:

$$y_0\int^{\xi} s^{-2}(u)du=\begin{cases} \dfrac{B(\xi)}{A_D W(D,B)\sqrt{n^*(\xi)}}+\text{cst} & \text{if} \quad A_D\neq 0 \quad \text{(C.6)}\\ \dfrac{D(\xi)}{A_B W(B,D)\sqrt{n^*(\xi)}}+\text{cst} & \text{if} \quad A_B\neq 0 \quad \text{(C.7)}\end{cases}$$

$$y_0\int^{\xi} s^{2}(u)du=A_B{}^2 W\left(B,\partial_\omega K|_0\right)+A_D{}^2 W\left(D,\partial_\omega P|_0\right)+A_B A_D\left[W\left(B,\partial_\omega P|_0\right)+W\left(D,\partial_\omega K|_0\right)\right]+\text{cst}\,, \tag{C.8}$$

where $A_B$ and $A_D$ are the two constant coefficients involved in the LC defining the general solution of eq. (16) when expressed for $n^{*\pm 1/2}$, instead of $\eta^{\pm 1/2}$, namely:

$$\begin{cases} A_B=\left(n_0^{*\pm 1/2}D_1-n_1^{*\pm 1/2}D_0\right)/\left(B_0D_1-B_1D_0\right)\\ A_D=\left(-n_0^{*\pm 1/2}B_1+n_1^{*\pm 1/2}B_0\right)/\left(B_0D_1-B_1D_0\right).\end{cases} \tag{C.9}$$

A sufficient condition for applying eq. (C.8) is that the linearly independent solutions $K(\xi,k_0)$ and $P(\xi,k_0)$ in eq. (21) be three times continuously differentiable in a neighborhood of $\xi$ =0 (for more details, see [36], [48], [49]).

Furthermore, for the $\langle E\rangle$-form profiles, by substituting eq. (C.6) or (C.7) into eq. (C.4) we get the following *implicit* relation:

$$z(\xi) = \frac{B_0 D(\xi) - B(\xi) D_0}{W(B,D)\sqrt{n_0^*}\sqrt{n^*(\xi)}} \tag{C.10}$$

An *explicit* relation can sometimes be obtained if it happens that a general relation of the following kind can be found between $B(\xi)$ and $D(\xi)$:

$$f_{BD}(B(\xi), D(\xi)) = \text{cst} \tag{C.11}$$

As a matter of fact, when extracting $B(\xi)$ and $D(\xi)$ from eq. (C.6) or (C.7) together with eq. (C.4), we get:

$$f_{BD}\left(\sqrt{n^*(z)}\left(-A_D W(B,D) z + B_0 n_0^{*-1/2}\right), \sqrt{n^*(z)}\left(A_B W(B,D) z + D_0 n_0^{*-1/2}\right)\right) = \text{cst}, \tag{C.12}$$

which, jointly with eq. (C.9), will provide an unique opportunity for getting an *explicit* definition of $n^*$ vs. $z$ as it is shown in some examples.

(2) TM-wave at oblique incidence

The integration of $2\int_0^\xi n^{*-1}(u)\left(1 \pm \sqrt{1 - 4I_a^2 n^{*-2}(u)}\right)^{-1} du$ is unlikely to lead to a closed-form analytic expression. We will have to resort to a numerical method for evaluating the relation between $\xi$ and $z$, which will be expressed through: $z = g_N(\xi)$. The index profile will then be defined *implicitly*:

$$\begin{cases} z = g_N(\xi) \\ n = \left(2n^{*-2}(\xi)\right)^{-1/2}\left(1 \pm \sqrt{1 - 4I_a^2 n^{*-2}(\xi)}\right)^{1/2} \end{cases}. \tag{C.13}$$

If it happens that the function $n^* = f(\xi)$ is analytically invertible, the implicit set will reduce to a *single implicit* relation:

$$z = g_N\left(f^{-1}\left(n^2 \big/ \sqrt{n^2 - I_a^2}\right)\right). \tag{C.14}$$

**ANNEX D**

The *joint PROperty & FIeld Darboux Transformations* (PROFIDT) method may be used for generating exact analytical solutions for modelling traveling waves in 1D-heterogeneous media, when the transport process can be described by a pair of coupled equations of the same form as eq. (2) and (3), that is:

$$\begin{cases} \dfrac{dF}{dz} = -i\omega g G & \text{(D.1)} \\ \dfrac{dG}{dz} = -i\omega f F\,, & \text{(D.2)} \end{cases}$$

where $f(z)$ and $g(z)$ are real-valued, of same sign and piecewise $C^1$ functions representing the variable parameters. The previous pair of coupled 1st order DE leads to two 2nd order ODE, namely linear wave equations with *variable-coefficients:*

$$\langle F \rangle: \qquad -\omega^2 F = f^{-1}\left(g^{-1} F_{,z}\right)_{,z} \tag{D.3}$$

$$\langle G \rangle: \qquad -\omega^2 G = g^{-1}\left(f^{-1} G_{,z}\right)_{,z} \tag{D.4}$$

The method is also suitable for problems described by only one among the latter 2nd order ODEs (yet it will give rise to solvable profiles of only one form instead of both $\langle F \rangle$-form and $\langle G \rangle$-form). The PROFIDT method gives the opportunity to construct solvable profiles of the property $s(\xi) = \eta(\xi)^{\pm 1/2}$, where $\eta = \sqrt{f/g}$, vs. the transformed independent-variable $\xi = \int_0^z \chi(u)du$ where $\chi = \sqrt{fg}$. Many examples can be found, in particular in [1], [2], [3]; let us just mention:

- lossless electrical transmission line with distributed inductance $L(z)$ and conductance $C(z)$ (lossless telegrapher's equation) [85], [86]. There is a correspondence between ($F, G, f, g$) and ($V, I, C, L$) where $V$ is the harmonic voltage and $I$ the intensity. The new variables $\chi$ and $\eta$ are defined by $\chi \equiv \sqrt{CL}$ and $\eta \equiv \sqrt{C/L}$ (admittance).
- longitudinal acoustic waves (as well as shear waves, in solids) in a graded medium with variable mass density $\rho(z)$ and velocity of sound $c(z)$ [1], [87], [88]. The correspondence is ($F, G, f, g$) $\leftrightarrow$ ($v, P, \rho, 1/\rho c^2$) where $v$ is the harmonic particle velocity or displacement and $P$ the acoustic pressure or stress; $\rho c^2$ is the elastic parameter (modulus). We get $\chi \equiv c^{-1}$ ($\xi$ corresponds to the integrated travel-time from 0 to the depth $z$) and $\eta \equiv \rho c$.
- ocean gravity waves over a beach with a variable water-height $H(z)$ [89]. The correspondence is ($F, G, f, g$) $\leftrightarrow$ ($v, h, h_a/(Hg_0), h_a/H^2$) where $v$ is the harmonic velocity in the horizontal direction, $h$ is the water height-fluctuation, $g_0$ is the gravitational acceleration and $h_a$ the reference height. In that case: $\chi \equiv \sqrt{h_a^2/(H^3 g_0)}$ and $\eta \equiv \sqrt{H/g_0}$.

Acknowledgment. This work was funded in the frame of the RGC-DSG Program: "Codes optiques". The author is grateful to Grégory Vincent, Riad Haïdar, Marion Krapez, Yves-Michel Frédéric and Grégoire Ky for valuable discussions.

| $V_{[0]}(\xi)=\beta$ | $j+1$ | $p_{j+1}$ | $\tilde{\psi}_{[j,p_{j+1}]}(\xi)\propto$ | $\sigma_{[j,p_{j+1}]}(\xi)$ | $V_{[j+1]}(\xi)$ | $\eta_{[j+1]}^{\pm 1/2}(\xi)\propto\begin{bmatrix}B_{[j+1]}\\ D_{[j+1]}\end{bmatrix}$ | nb. of param. | $\psi_{[j+1]}(\xi,k_0)\propto\begin{bmatrix}K_{[j+1]}\\ P_{[j+1]}\end{bmatrix}$ | Case |
|---|---|---|---|---|---|---|---|---|---|
| $\beta=0$ | 0 | - | - | - | 0 | $\begin{bmatrix}1\\ \xi\end{bmatrix}$ | 2 | $\begin{bmatrix}C^{(0)}\\ S^{(0)}\end{bmatrix}$ | c1 |
| $\beta=0$ | 1 | 0 | $\xi+\gamma$ | $(\xi+\gamma)^{-1}$ | $2(\xi+\gamma)^{-2}$ | $\begin{bmatrix}(\xi+\gamma)^{-1}\\ (\xi+\gamma)^{2}\end{bmatrix}$ | 3 | $\begin{bmatrix}(ik_0-\sigma)C^{(0)}\\ (-ik_0-\sigma)S^{(0)}\end{bmatrix}$ | c2 |
| $\beta=0$ | 1 | $p_1>0$ | $\begin{bmatrix}\cosh(\xi/\xi_c)\\ \sinh(\xi/\xi_c)\end{bmatrix}$ | $\xi_c^{-1}\tanh(\hat{\xi})$ | $-2\xi_c^{-2}\cosh^{-2}(\hat{\xi})$ | $\begin{bmatrix}\tanh(\hat{\xi})\\ 1-\hat{\xi}\tanh(\hat{\xi})\end{bmatrix}$ | 4 | $\begin{bmatrix}(ik_0-\sigma)C^{(0)}\\ (-ik_0-\sigma)S^{(0)}\end{bmatrix}$ | c3 |
| $\beta=0$ | 1 | $p_1>0$ | $\begin{bmatrix}\cosh(\xi/\xi_c)\\ \sinh(\xi/\xi_c)\end{bmatrix}$ | $\xi_c^{-1}\coth(\hat{\xi})$ | $2\xi_c^{-2}\sinh^{-2}(\hat{\xi})$ | $\begin{bmatrix}\coth(\hat{\xi})\\ 1-\hat{\xi}\coth(\hat{\xi})\end{bmatrix}$ | 4 | $\begin{bmatrix}(ik_0-\sigma)C^{(0)}\\ (-ik_0-\sigma)S^{(0)}\end{bmatrix}$ | c4 |
| $\beta=0$ | 1 | $p_1<0$ | $\begin{bmatrix}\cos(\xi/\xi_c)\\ \sin(\xi/\xi_c)\end{bmatrix}$ | $-\xi_c^{-1}\tan(\hat{\xi})$ | $2\xi_c^{-2}\sin^{-2}(\hat{\xi})$ | $\begin{bmatrix}\tan(\hat{\xi})\\ 1+\hat{\xi}\tan(\hat{\xi})\end{bmatrix}$ | 4 | $\begin{bmatrix}(ik_0-\sigma)C^{(0)}\\ (-ik_0-\sigma)S^{(0)}\end{bmatrix}$ | c5 |
| $\beta>0$ | 0 | - | - | - | $\beta$ | $\begin{bmatrix}\cosh(\xi/\xi_c)\\ \sinh(\xi/\xi_c)\end{bmatrix}$ | 3 | $\begin{bmatrix}C^{(\beta)}\\ S^{(\beta)}\end{bmatrix}$ | c6 |
| $\beta>0$ | 1 | 0 | $\begin{bmatrix}\cosh(\xi/\xi_c)\\ \sinh(\xi/\xi_c)\end{bmatrix}$ | $\xi_c^{-1}\tanh(\hat{\xi})$ | $\xi_c^{-2}\left(-2\cosh^{-2}(\hat{\xi})+1\right)$ | $\begin{bmatrix}\operatorname{sech}(\hat{\xi})\\ \sinh(\hat{\xi})+\hat{\xi}\operatorname{sech}(\hat{\xi})\end{bmatrix}$ | 4 | $\begin{bmatrix}\left(i\sqrt{k_0^{2}-\beta}-\sigma\right)C^{(\beta)}\\ \left(-i\sqrt{k_0^{2}-\beta}-\sigma\right)S^{(\beta)}\end{bmatrix}$ | c7 |
| $\beta>0$ | 1 | 0 | $\begin{bmatrix}\cosh(\xi/\xi_c)\\ \sinh(\xi/\xi_c)\end{bmatrix}$ | $\xi_c^{-1}\coth(\hat{\xi})$ | $\xi_c^{-2}\left(2\sinh^{-2}(\hat{\xi})+1\right)$ | $\begin{bmatrix}\operatorname{csch}(\hat{\xi})\\ \cosh(\hat{\xi})-\hat{\xi}\operatorname{csch}(\hat{\xi})\end{bmatrix}$ | 4 | $\begin{bmatrix}\left(i\sqrt{k_0^{2}-\beta}-\sigma\right)C^{(\beta)}\\ \left(-i\sqrt{k_0^{2}-\beta}-\sigma\right)S^{(\beta)}\end{bmatrix}$ | c8 |
| $\beta<0$ | 0 | - | - | - | $\beta$ | $\begin{bmatrix}\cos(\xi/\xi_c)\\ \sin(\xi/\xi_c)\end{bmatrix}$ | 3 | $\begin{bmatrix}C^{(\beta)}\\ S^{(\beta)}\end{bmatrix}$ | c9 |
| $\beta<0$ | 1 | 0 | $\begin{bmatrix}\cos(\xi/\xi_c)\\ \sin(\xi/\xi_c)\end{bmatrix}$ | $-\xi_c^{-1}\tan(\hat{\xi})$ | $\xi_c^{-2}\left(2\sin^{-2}(\hat{\xi})-1\right)$ | $\begin{bmatrix}\sec(\hat{\xi})\\ \sin(\hat{\xi})+\hat{\xi}\sec(\hat{\xi})\end{bmatrix}$ | 4 | $\begin{bmatrix}\left(i\sqrt{k_0^{2}-\beta}-\sigma\right)C^{(\beta)}\\ \left(-i\sqrt{k_0^{2}-\beta}-\sigma\right)S^{(\beta)}\end{bmatrix}$ | c10 |

- For case c2: $\gamma$ is a free-parameter in $]-\infty,-\xi_1[\,\cup\,]0,+\infty[$.
- For cases c3-c5, c7, c8 and c10, $\hat{\xi}$ is a transformed optical-thickness defined by $\hat{\xi}(\xi)\equiv\xi/\xi_c+\gamma$ where $\gamma$ is a free-parameter in $]-\infty,+\infty[$.
- For cases c3-c5: $\xi_c\equiv|p_1|^{-1/2}$; for cases c6-c10: $\xi_c\equiv|\beta|^{-1/2}$; $\xi_c$ corresponds to the *characteristic* optical-thickness of the profile element.

- In col.9, the symbol $\sigma$ is used for $\sigma_{[j,p_{j+1}]}$ which should be taken from col. 5 ; $C^{(\beta)}$ and $S^{(\beta)}$ represent the functions $\exp\left(i\sqrt{k_0^{\ 2}-\beta}\xi\right)$ and $\exp\left(-i\sqrt{k_0^{\ 2}-\beta}\xi\right)$.

Table 1: Main results before and after the application of one PROFIDT sequence when starting from a constant potential $V_{[0]}(\xi)=\beta$. In column 1-to 10 we find successively: the seed potential $\beta$, the order $j+1$ of the Darboux Transformation, the currently chosen transformation parameter $p_{j+1}$, the currently chosen transformation function $\tilde{\psi}_{[j,p_{j+1}]}(\xi)$, the logarithmic derivative $\sigma_{[j,p_{j+1}]}(\xi)$ (eq. (A.2)), the transformed potential $V_{[j+1]}(\xi)$ (eq. (A.4)), the LC defining the new admittance profile of $\langle E\rangle$-form and $\langle H\rangle$-form through $\eta_{[j+1]}^{\ \pm 1/2}(\xi)$ (eq. (A.7)), the number of free-parameters defining these profiles, the LC defining the new EM solution $\psi_{[j+1]}(\xi,k_0)$ (eq. (A.5)), and the case number. For the sake of completeness, the fundamental solutions are also reported here, namely cases c1, c6, c9 for, respectively, a zero potential, a constant positive potential, and a constant negative potential (adapted from Table 1 in [36]).

| Inferred variable from $\eta_r$ | TE-mode | TM-mode |
|---|---|---|
| $n^*(\xi)$ | $n^*=\eta_r$ | |
| $\varepsilon_r(\xi)$ | $\varepsilon_r=\eta_r^{\ 2}+I_a^{\ 2}$ | $\varepsilon_r=\left(2\eta_r^{-2}\right)^{-1}\left(1\pm\sqrt{1-4I_a^2\eta_r^{-2}}\right)$ |
| $n(\xi)$ | $n=\left(\eta_r^{\ 2}+I_a^{\ 2}\right)^{1/2}$ | $n=\left(2\eta_r^{-2}\right)^{-1/2}\left(1\pm\sqrt{1-4I_a^2\eta_r^{-2}}\right)^{1/2}$ |

Table 2: Correspondence between the relative admittance profile $\eta_r(\xi)$ and the inferred profiles of pseudo-index $n^*(\xi)$, permittivity $\varepsilon_r(\xi)$ and refractive index $n(\xi)$ in the case of *non-magnetic* materials ($\mu_r(\xi)$=1).